\documentclass[preprint2]{aastex631wtr}

\def\bfc{}

\def\kms{km~s$^{-1}$}
\def\cm3{cm$^{-3}$}
\def\apj{ApJ}
\def\apjl{ApJL}

\def\aap{A\&A}

\usepackage{graphicx}
\usepackage{epsf}
\usepackage{natbib}

\newcommand{\simgt}{\ga}
\newcommand{\simlt}{\lower.5ex\hbox{$\; \buildrel < \over \sim \;$}}

\begin{document}

\title{Supernova Shocks in Molecular Clouds: Shocks Driven into Dense Cores in IC~443 and 3C~391}

\shorttitle{Supernova Shocks in Prestellar Cores }

\author[0000-0001-8362-4094]{William T. Reach}
\affil{Space Science Institute, 4765 Walnut St, Suite B, Boulder, CO 80301, USA}
\email{wreach@spacesciece.org}

\author[0000-0002-6488-8227]{Le Ngoc Tram}
\affil{Max-Planck-Institut f\"ur Radioastronomie, Auf dem H\"ugel 69, D-53121, Bonn, German}
\affil{Leiden Observatory, Leiden University, PO Box 9513, 2300 RA Leiden, The Netherlands}

\author[0000-0002-6528-3836]{Curtis DeWitt}
\affil{Space Science Institute, 4765 Walnut St, Suite B, Boulder, CO 80301, USA}

\author[0000-0002-8447-4398]{Pierre Lesaffre}
\affil{Laboratoire de Physique de l'\'Ecole Normale Sup\'erieure, ENS, Universit\'e PSL, CNRS, Sorbonne Universit\'e, Universit\'e Paris Cit\'e, 75005, Paris, France}

\author[0000-0003-0060-8887]{Benjamin Godard}
\affil{Laboratoire de Physique de l'\'Ecole Normale Sup\'erieure, ENS, Universit\'e PSL, CNRS, Sorbonne Universit\'e, Universit\'e Paris Cit\'e, 75005, Paris, France}

\def\norespond{
\author[0000-0002-8594-2122]{Matthew Richter}
\affil{Department of Physics,    University of California, Davis,    Davis, CA 95616, USA}
}

\author[0000-0002-0354-1684]{Antoine Gusdorf}
\affil{Laboratoire de Physique de l'\'Ecole Normale Sup\'erieure, ENS, Universit\'e PSL, CNRS, Sorbonne Universit\'e, Universit\'e Paris Cit\'e, 75005, Paris, France}
\affil{Observatoire de Paris, PSL University, Sorbonne Université, LERMA, 75014 Paris, France}

\begin{abstract}
Supernova shocks into dense molecular cores in IC 443 (clumps B, C, and G) and 3C 391 were observed using the Stratospheric Observatory for Infrared Astronomy and complemented by archival data from the
{\it Herschel} Space Observatory.
The pure rotational transitions 0-0 S(1) and S(5) of H$_2$, and the ground-state  $1_{10}-1{01}$ transition 
of H$_2$O, are all broadened, arising from molecules that survive the passage of the shock front. 
Theoretical models
from the Paris-Durham shock code were analyzed to generate synthetic profiles that approximately match
the observations. {\bfc The observations can be fit with two shock conditions,
which approximate the range of densities in the pre-shock molecular cloud.}
The width and brightness of the S(5) lines require shocks into gas with a
density of order 2,000 cm$^{-3}$, into which the IC 443 blast wave drives shocks
with speed 60 \kms. 
The brightness and narrower width of the S(1) lines requires different shocks, into 
gas with density of order $10^5$ cm$^{-3}$, with shock speeds of 10 \kms.
The H$_2$O velocity distribution is also consistent with these shocks.
The existence of shocks into dense gas shows that the bright shocked clumps
in IC~443 were prestellar cores. 
It is unlikely that they will form stars soon after the passage of the shock front, 
given the input of kinetic and thermal energy from the shocks.
\end{abstract}

\section{Introduction}

The two supernova remnants, IC~443 and 3C~391, present two of the most accessible sites  of
supernova-shocked molecular gas.
The southern portion of  IC~443 is impacting a molecular cloud along an
extensive and clumpy ridge of shocked molecular gas first seen in
mm-wave CO emission \citep{denoyer1979,huang1986} and
near-infrared rovibrational 1-0 S(1) emission
\citep{burton1988}.
Submillimeter lines reveal a wealth of chemistry as well as dynamic changes across the shock ridge
\citep{vandishoeck1993}.
A pure rotational line of H$_2$, 0-0 S(4),  expected to be a strong coolant for many types
of molecular shock, has been detected in ground-based observations
\citep{richter1995}, but lines from the lowest energy levels are inaccessible at ground level.
The Infrared Space Observatory (ISO) detected S(0) through S(8) lines from the molecular ridge
in IC 443, and the results indicated young (1,000 to 2,000 yr) shocks into moderate density 
($10^4$ cm$^{-3}$) gas \citep{cesarsky1999}.
{\it Spitzer} detected and imaged H$_2$ emission lines \citep{neufeld2007,yuan2011,dellova2024}.
The Submillimeter-Wave Astronomical Satellite (SWAS) detected H$_2$O emission from the shocked gas,
with a remarkably wide velocity distribution
\citep{snell2005}.
The supernova remnant 3C~391 impinges upon a giant molecular cloud
\citep{wilner1998}, and shocked clumps are evident in 
infrared  \citep{reach1996,reach1998}
mm-wave \citep{reach1999,neufeld2014}. 

\begin{figure}
    \centering
    \includegraphics[width=0.5\textwidth]{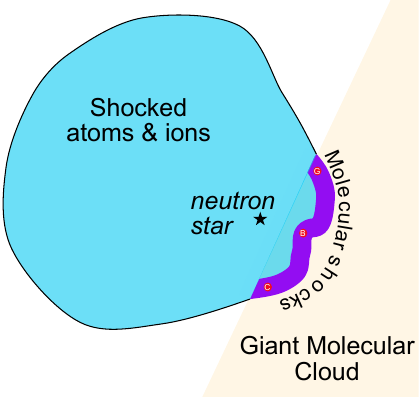}
    \caption{Cartoon showing the context for supernova remnants IC~443 and 3C~391, produced by supernovae near or just inside the edges of molecular clouds. {\bfc The blast wave propagates through lower-density gas making X-rays (blue), then into the molecular clumps (purple) and dense
    cores (red, labeled for those of IC~443).}}
    \label{fig:blowout}
\end{figure}

Figure~\ref{fig:blowout} summarizes the context for IC~443 and 3C~391.
The stellar remnant for IC~443 has been found \citep{olbert2001}, but none has yet been found for 3C~391. The explosion sites were likely
closer to the surfaces of their molecular clouds than the geometric centers of the X-ray emission or the radio shell
that include much faster shocks into lower-density gas. The ages and association with molecular clouds are similar
for the two supernova remnants.
A significant difference is their distance:  $\sim$2 kpc for IC~443, $\sim$7 kpc for 3C 391 \citep{lee2020}. Both supernova remnants are very radio bright \citep{green2019}; IC~443 is brighter as seen from Earth, but 3C~391 is actually more radio luminous.

The IC~443 and 3C~391 supernova blast waves are impinging upon dense ($10^3$ to $10^5$ cm$^{-3}$) molecular gas, 
providing a detailed view of shock fronts where the parent molecular clouds of the progenitors are being disrupted.
The warm, shocked gas and compressed magnetic field are also an apparent origin for cosmic rays,
which make IC~443 a prominent $\gamma$-ray source \citep{abdo2010}.
The $\gamma$-rays from IC~443 have been interpreted as interaction between cosmic rays and a molecular cloud
in front of the supernova remnant \citep{torres2008}, where fast ($\sim 150$ \kms) shock fronts act as a `hadronic accelerator'
\citep{tavani2010}.

In this paper, we study the emission from molecular hydrogen and water using new observations of
the velocity-resolved spectral line profiles of transitions among low energy levels.
We concentrate on two H$_2$ lines that arise from very different energy levels.
The first part of this project \citep[][hereafter referred to as Paper 1]{reach2019} was to measure the 0-0 S(5) line of H$_2$, 
a ground vibrational state transition of ortho-H$_2$ between the rotational J=7 and J=5 levels. The S(5) line is the brightest line
emerging from a wide range of molecular
shocks with the pressure of the blast wave of a $10^4$ year old supernova remnant. (The age and pressure
of the blast wave are derived in Paper 1.) 
The new observational data in this paper are
velocity-resolved spectra of the S(1) line, which we combine with data from Paper 1 and archival observations of a ground-state H$_2$O
line to  better constrain 
the nature of the shock fronts that supernova remnants drive into molecular clouds. 

\section{Observations\label{sec:obs}}

\subsection{SOFIA}
Observations were made using the Stratospheric Observatory for Infrared Astronomy (SOFIA), which
was an infrared-optimized telescope on a 747 aircraft, operated by the United States and German
space agencies \citep{youngEarlyScienceSOFIA2012}
through 2022. The infrared spectra were taken with the Echelon-Cross-Echelle Spectrograph \citep[EXES,][]{richter2018},
with a resolution power of 50,000 (6 \kms). 
The spectral configuration was centered on the 587 cm$^{1}$ (17.03 $\mu$m) pure rotational ground state transition
of ortho molecular hydrogen, i.e. the 0-0 S(1) line.
The 3.2$''$ slit was centered on a known location of bright H$_2$
emission for each of four supernova remnant molecular shocks. 

The observations of IC~443 focus on the well-known shocked clumps B, C, and G \citep{huang1986}.
Each of these shocked regions is large (up to an arcmin), so the precise observing locations were derived
from more recent imaging observations.
For IC~443~C, we observed the peak in
the H$_2$ S(5) emission line map from {\it Spitzer} \citep{neufeld2007}.
For the other three sources, we observed the same positions as were previously observed in the S(5) line with SOFIA previously  (Paper 1).
Specifically, we observed a 2MASS K$_s$ peak (also used in Paper 1) due to vibrationally excited H$_2$ in IC~443~G;
the {\it Herschel} H$_2$O and [O I] peak in IC~443~B (from archival observation 1342227773);
and a SOFIA H$_2$ S(5) peak in 3C~391 from Paper 1.

\def\tnm{\tablenotemark}
\begin{deluxetable}{ccccc}
\tablecolumns{5}
\tabletypesize{\scriptsize}
\tablecaption{SOFIA Observing Log\tablenotemark{a}\label{obstab}} 
\tablehead{
\colhead{Date} & \colhead{Flight} & \colhead{Altitude} & \colhead{Elevation} & \colhead{Time} \\
\colhead{} & \colhead{} &  \colhead{(feet)} & \colhead{($^\circ$)} & \colhead{(min)}
}
\startdata
\sidehead{IC~443 C (06:17:42.4 +22:21:27)}
2022 Apr 27 & 860 & 38,020 & 37 & 22  \\
2022 Apr 28 & 861 & 39,000 & 30 & 12  \\
2022 Apr 29 & 862 & 39,010 & 25 & 10  \\
\hline
\sidehead{IC~443 G (06:16:43.6 +22:32:37)}
2022 Apr 27 & 860 & 39,000 & 26 & 6  \\
2022 Apr 28 & 861 & 39,010 & 36 & 13  \\
\hline
\sidehead{IC~443 B (06:17:16.1 +22:25:15): }
2022 Apr 29 & 862 & 39,010 & 35 & 138  \\
\hline
\sidehead{3C~391 (18:49:22.6 -00:57:17)}
2022 Apr 27 & 860 & 43,020 & 53 & 20 \\
\enddata
\tablenotetext{a}{target names and J2000 coordinates, with observing conditions for each flight on which they were observed}
\end{deluxetable}

\begin{figure*}
\includegraphics[width=\textwidth]{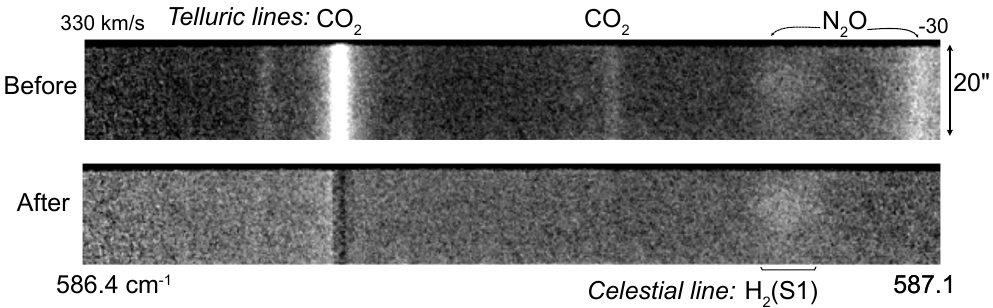}
\caption{Slit spectrum of  the H$_2$ S(1) line from IC~443~B showing subtraction of residual sky emission. The upper panel shows one  echellle order  from standard nod subtraction. Sky lines from CO$_2$ and N$_2$O remain, with the N$_2$O lines partially overlapping the region of celestial H$_2$ S(1) emission. The lower panel shows the same order, after subtracting an additional small fraction of the sky spectrum. The H$_2$ line is labeled below the lower panel. While the H$_2$ line is visible in both panels, the suppression of sky features significantly improves the accuracy of measuring the line profile. The spectrum used for analysis was extracted from an $8''$ portion of the slit near its center.}
\label{fig:subtraction}
\end{figure*}

Table~\ref{obstab} summarizes the observing conditions, with the aircraft
altitude and telescope elevation specified because telluric absorption depends significantly on those
two factors. 
The telescope was nodded between the target position and a nearby position with no trace of H$-2$ in the 2MASS K$_s$ band or {\it Spitzer} line images. But the sky subtraction is never perfect, leading to contamination of the 
nod-differenced spectra by emission lines from the Earth's atmosphere. 
We devised a method to remove the problematic residual emission by adding and subtracting small fractions of the sky radiance spectrum (the sky flux before nod subtraction). 
In all H$_2$ S(1) observations there is only ever one order in the echellogram containing the 
587 cm$^{-1}$ H$_2$ emission line, even considering the various Doppler shifts to the different targets. 
At most, two atmospheric lines, from N$_2$O and CO$_2$ contaminate the celestial H$_2$ emission.
The sky radiance was scaled by a factor ($\pm 0.5$\%) that matches the amplitude of the features 
observed in the nod difference, then it was subtracted. Figure~\ref{fig:subtraction} illustrates the 
before and after results in the two-dimensional echellogram for IC~443~B.

\begin{figure*}
\includegraphics[width=\textwidth]{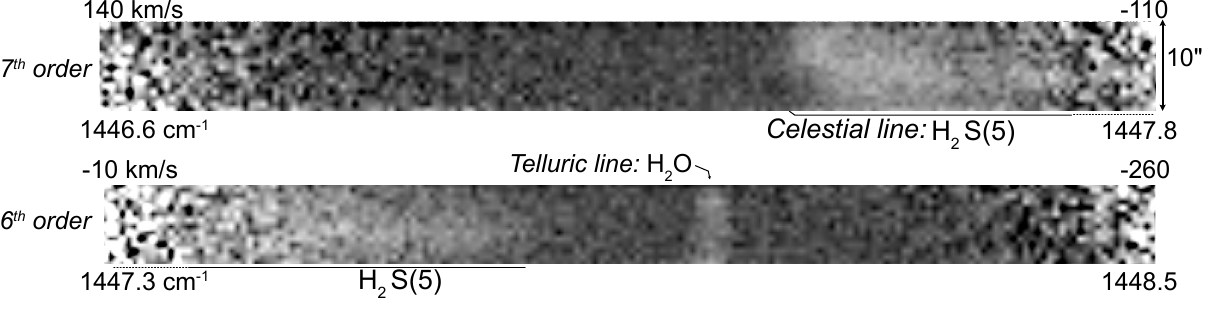}
\caption{Spectral image for the IC~443~B observations of the H$_2$ S(5) line. Two orders of the echelle spectrum 
are shown, because they both cover part of the very broad celestial H$_2$ line. 
The minimum and maximum wavenumber and velocity of each order are labeled on their lower and upper corners, respectively. An $8''$ portion of the slit, toward the top, was extracted to match
the region extracted for the S(1) line. 
}
\label{fig:s5image}
\end{figure*}

The S(5) line observations (from Paper 1) were not as significantly affected by atmospheric effects. Figure~\ref{fig:s5image} shows
the echelle spectral image for two orders that include part of the velocity range where IC~443~B shocked gas is detected.
No telluric features are visible on this scale. There is some evidence for a change in the velocity distribution across
the $10''$ slit, with the upper part of the slit somewhat more concentrated to lower velocities, and the lower part of the slit having a broader velocity distribution. 

We used $8''$ portions of the S(1) and S(5) observed slits to extract the spectra used for analysis in this paper. 
During the observation,
the center of the slit for the S(1) observation was adjusted to the peak of S(5) emission, which is why the H$_2$ S(1) emission is centered in the slit in Figure~\ref{fig:subtraction} but the S(5) emission is near the top of the slit in Figure~\ref{fig:s5image}.

\begin{figure*}
\gridline{
    \fig{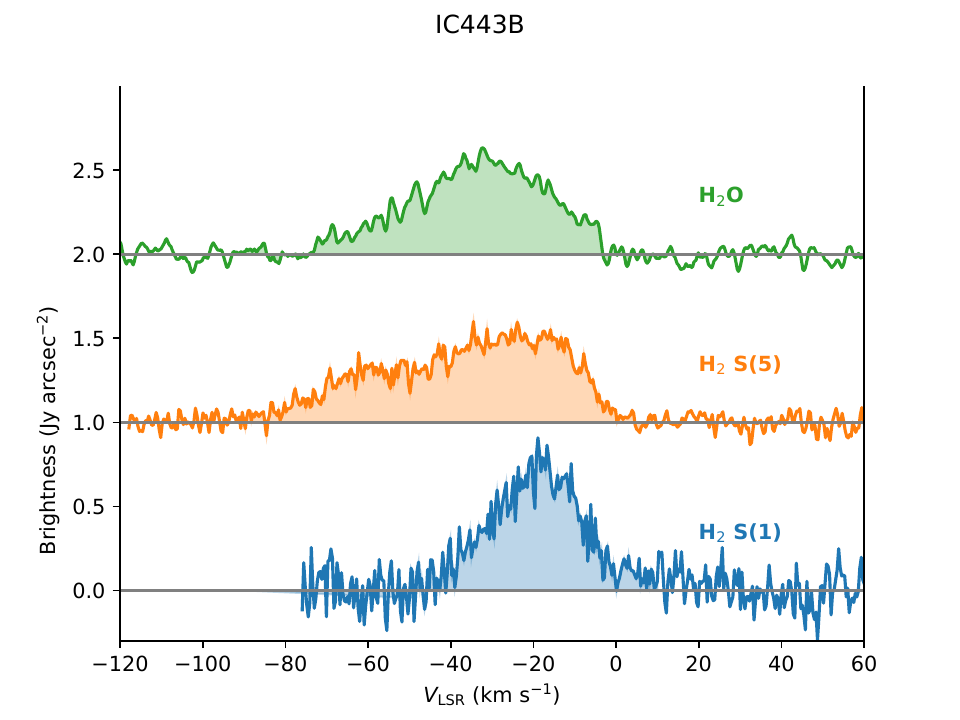}{0.49\textwidth}{}
    \fig{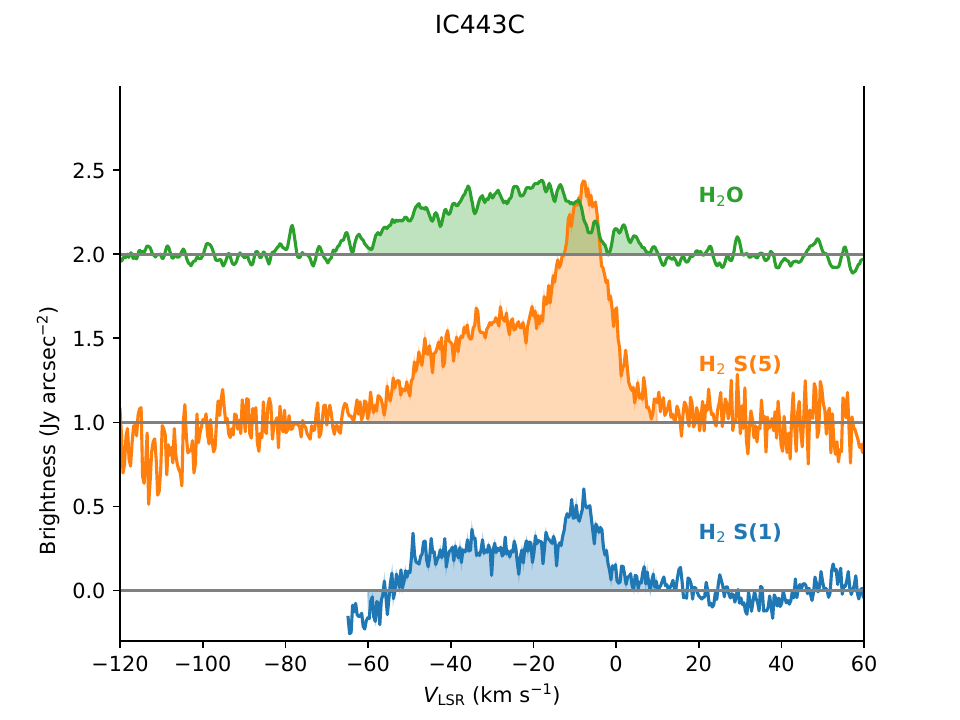}{0.49\textwidth}{}
    }
\gridline{
    \fig{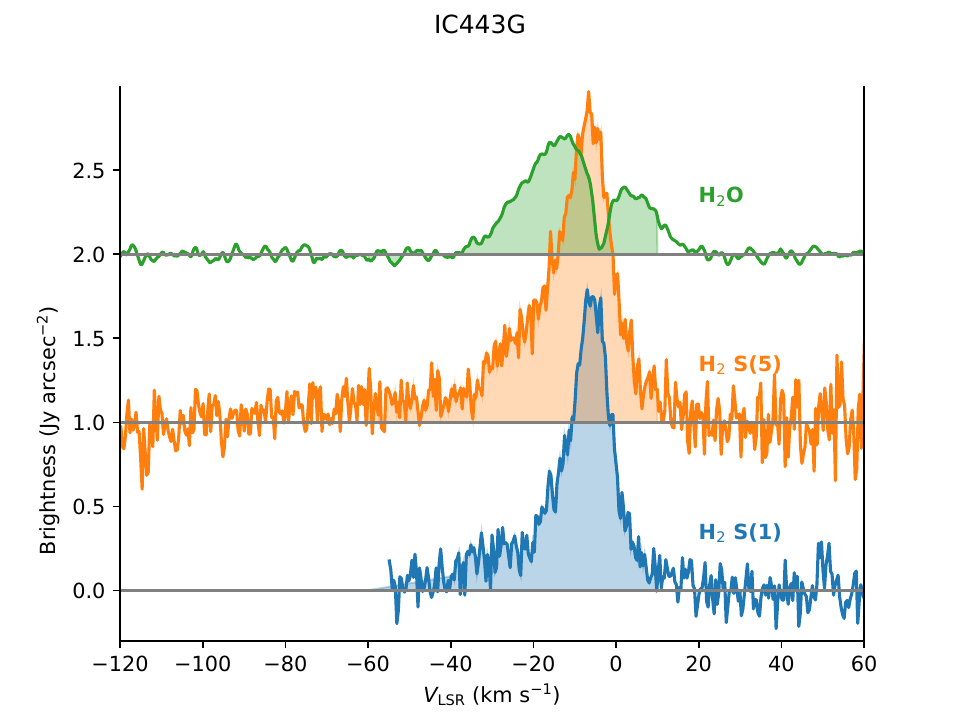}{0.49\textwidth}{}
    \fig{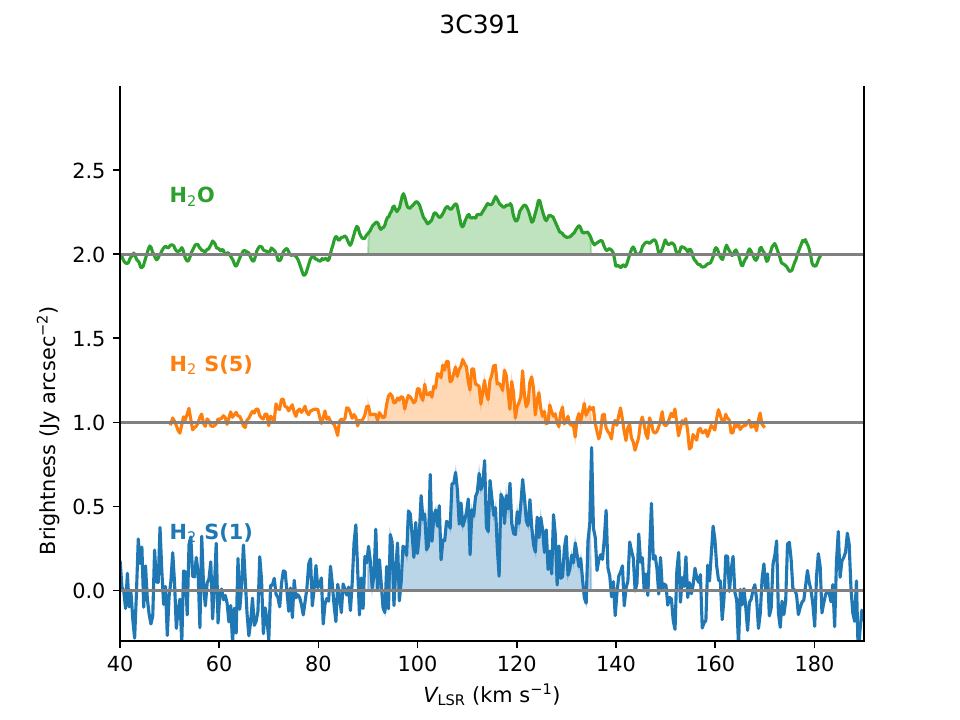}{0.49\textwidth}{}
    }
\caption{SOFIA spectra of H$_2$ S(1) and S(5) and H$_2$O $1_{10}-1_{01}$ lines from IC 443 clumps B, C, G,  and from 3C 391
at the positions listed in Table~\ref{obstab}.
The H$_2$ lines are in units of Jy~arcsec$^{-2}$ and the H$_2$O line is in units of brightness temperature (K).
The S(5) and H$_2$O spectra are vertically offset by 1 and 2 for clarity. All brightnesses for 3C~391 were doubled to be on the same scale.
}\label{fig:spectra}
\end{figure*}

Figure~\ref{fig:spectra} shows the S(1) and S(5) profiles.
The S(5) spectrum is significantly different from S(1), in particular for IC~443~B. To determine whether there are strong localized variations, subslit
spectra were extracted from the top and bottom half of the S(5) slit for IC~443~B. There is in fact a difference, but only in the amount of lower-velocity emission, between -17 and 0 \kms. Thus it appears that the high-velocity wing (-20 to -60 \kms) is widespread across the $10''$ slit, and the lower-velocity emission is
more spatially concentrated.
The S(1) emission concentrated to lower velocities and the center of the slit.

\begin{deluxetable*}{ccccccccccccc}
\tablecolumns{12}
\tablecaption{Properties of the observed spectral lines \label{fittab}} 
\tablehead{
\multicolumn{1}{c}{Source} && \multicolumn{3}{c}{H$_2$ S(1) } && \multicolumn{3}{c}{H$_2$ S(5) } && 
\multicolumn{3}{c}{H$_2$O 1$_{10}$-$1_{01}$ }\\
\cline{1-1}
\cline{3-5}
\cline{7-9}
\cline{11-13}
\colhead{model}  &&
\colhead{$\langle v \rangle$} & \colhead{FWHM} & \colhead{$I_{\rm peak}$} &&
\colhead{$\langle v \rangle$} & \colhead{FWHM} & \colhead{$I_{\rm peak}$} &&
\colhead{$\langle v \rangle$} & \colhead{FWHM}  & \colhead{$I_{\rm peak}$}\\ 
 && \colhead{(\kms)} & \colhead{(\kms)} & \colhead{(Jy~arcsec$^{-2}$)} &&
\colhead{(\kms)} & \colhead{(\kms)} & \colhead{(Jy~arcsec$^{-2}$)} &&
\colhead{(\kms)} & \colhead{(\kms)} & \colhead{(Jy~arcsec$^{-2}$)}
}
\startdata
IC 443 B && -19.6 & 28.6 & 0.91 && -36.6 & 48.0 & 0.6 && -34.3 & 36.7 & 0.06\\
IC 443 C && -20.7 & 33.5 & 0.60 && -19.9 & 37.1 & 1.4 && -27.2 & 37.0 & 0.04\\
IC 443 G && -9.9  & 23.2 & 1.8  && -11.2 & 24.0 & 2.0 && -11.4 & 25.0 & 0.12\\
3C 391   && 111.4 & 24.1 & 0.39 && 110.0 & 21.8 & 0.4 && 111.2 & 28.1 & 0.03 
\enddata
\end{deluxetable*}

\subsection{Archival}

Observations from the ESA {\it Herschel} Space Observatory \citep{pilbratt2010} taken with the
HIFI \citep{degraauw2010} instrument were obtained from the ESA science archive.
The specific observation identifiers were 11342205310 (IC~443~C), 11342205311 (IC~443~B), and 11342205312 (IC~443~G), 
and 1342218432 (3C~391).
Each observation was an on-the-fly spectral map, with the receiver tuned to the H$_2$O 1$_{10}$-1$_{01}$ spectral line
at 556.936 GHz in the lower sideband of the dual-sideband system. 
The spectral resolution was 0.6 km~s$^{-1}$, and the diffraction resolution was $38''$.
The pixel in the HIFI spectral cube containing the SOFIA-observed position was extracted,
and the horizontal and vertical polarizations were averaged.

Figure~\ref{fig:spectra} shows the H$_2$O spectra. The velocity distribution is generally  
similar to that of the H$_2$ S(5) line, except for IC~443~G, where there is a deep absorption dip at -4 \kms (discussed in \S\ref{sec:h2oabsorption} below).

\def\extra{
Observations from the
Submillimeter Wave Astronomy Satellite \citep[(SWAS)][]{melnick2000}
were obtained from the NASA/IPAC/Caltech Infrared Science Archive (IRSA).
of the H$_2$O 1$_{10}$-1$_{01}$ line and the \ion{C}{1} 492 GHz fine-structure line.
The angular resolution was $200''\times 280''$.
The H$_2$O spectra for IC 443 were analyzed by \citet{snell2005}. 
We used the SWAS H$_2$O data as an initial reference and link with the previous work, but
the {\it Herschel} spectra have angular resolution significantly higher than SWAS.
(Even the {\it Herschel} spectra, however have lower angular resolution than the $\sim 2''$ resolution of the SOFIA H$_2$ observations.)
}


Table~\ref{fittab} summarizes the observed properties of the spectral lines. The line centroid and FWHM 
were measured using the first and second velocity moments of the velocity spectra, respectively.
The peak intensity is in the observational units of Jy~arcsec$^{-2}$, where 1 Jy is $10^{-23}$ 
erg~s~cm$^{-2}$~Hz$^{-1}$.

\section{Modeling the Supernova Shocks into Molecular Cores\label{sec:mods}}

Theoretical models of shocks into dense clumps 
were calculated with the Paris-Durham shock code 
\citep{flower2015,lesaffre2013,godard2019}.
The models are similar to those in the extensive grid \citep{kristensen2023}
that is
available from the interstellar medium database 
(ISMDB\footnote{\url{http://ism.obspm.fr/}}) of the Paris Observatory.
In summary, the models begin with interstellar conditions calculated 
for at equilibrium for a specific ($n_0$), external radiation field ($\chi$),
cosmic ray ionization rate $\zeta_{\rm CR}$, and magnetic field $B$. 
The output from the static calculation is used to initialize a photodissociation region (PDR),
which is slowly advected to the column density $2\times 10^{20}$ cm$^{-2}$, to calculate the pre-shock
abundances of all species. 
The models considered in this paper all have radiation field equal to the average interstellar value ($\chi=1$)
based on \citet{mathisInterstellarRadiationField1983}, with an intervening extinction $A_{\rm V}$=1.
The magnetic field strength is set for each pre-shock density according to flux freezing scaling, with a 
normalization such that $B=\beta  n_0^{-0.5}$ $\mu$G
(with $n_0$ in cm$^{-3}$). 

The cosmic ray ionization rate was set to $\zeta_{\rm CR}=5\times 10^{-16}$ s$^{-1}$,
which is $17\times$ higher than the models presented in the initial study of irradiated shocks \citep{godard2019},
inspired by the high cosmic ray ionization rate inferred from the H$_3^+$ abundance \citep{indriolo2010} and
the detection of TeV $\gamma$-rays toward IC 443 \citep{tavani2010,sinitsyna2023}.
Models with $\zeta$ a factor of 5 higher and lower than this were examined for differences,
but they were not significant for H$_2$ lines that are the primary subject of this paper.

The physics of shocks into dense, cold gas is largely determined by the ratio of the shock speed to the Alfv\'en speed in the charged
fluid taking into account charged dust grains:
\begin{equation}
V_{\rm A,c} = \frac{B}{\sqrt{4\pi \rho_{\rm c}}} \simeq 21 \beta \, {\rm km~s}^{-1}
\end{equation}
for the estimated abundance and charge of small grains in molecular clouds
\citep[as in][]{guillet2007}.
Shocks with speed $V_s > V_{\rm A,c}$ have an abrupt `jump' in physical conditions at the shock front, followed
by gradual cooling; these are commonly referred to as \textbf{J-type}. Molecules can be destroyed in fast shocks, and by the time
they reform, the gas velocities are low. 
The slower, \textbf{C-type} shocks occur at $V_s < V_{\rm A,c}$
because the neutral species decouple from the ions, which remain strongly coupled to the magnetic field \citep{draine1993}. 
This allows neutrals to flow past the initial shock front and avoid the abrupt heating that ions experience,
allowing some molecules (notably H$_2$) to avoid dissociation.

The velocities at which molecular emission arises depend on the shock type and velocity. In this discussion and throughout this paper, 
velocities will be measured in the rest frame of the observer 
\citep[`local standard of rest';][]{binney1998}. This is
important because it is shared by all gas struck by the blast wave: {\bfc we assume that all gas starts at rest and is impacted by the blast wave at 
the same time}.
Other frames, such as the commonly used one comoving with the shock front, depend upon the shock velocity, which is
not known {\it a apriori} and differs for gas of different density, making it difficult to compare to observational data.
Let's call the pre-shock velocity $V_0$ (with respect to the observer), with temperature $T_0$, and the shock velocity $V_s$ 
(with respect to the gas being shocked). The  peak temperature
$T_s=53 V_s^2$ K for a J-type shock, with $V_s$ in km~s$^{-1}$ \citep{lesaffre2013}.
For a J-type shock, a very thin layer of hot molecules ($T\sim T_s$) can exist at the shock front, but the bulk
of the molecules only return after molecules reform in the cooled ($T\sim T_0$) and decelerated gas; see \citet{lehmann2020} for a modern description of J-type shocks in dense gas.
The observed emission from a dissociative J-type shock would comprise highly-excited molecules at velocity $V_s+V_0$ and cold (reformed) molecules at $\sim V_0$, with
narrow (thermal) velocity distributions. Emission from the colder, slower layer would be  from lower energy levels.
For a C-type shock, the molecules persist throughout the post-shock gas, and molecular emission can arise from velocities ranging from 
$V_s+V_0$ to $V_0$. The shape of the velocity distribution
depends upon the  cooling and molecule reformation rate in the post-shock gas. 
For very young shocks, a CJ-type shock regime exists \citep{lesaffre2004}, for which the
region very close to the shock front is treated as a J-type shock and the larger-scale evolution
as a C-shock. Molecular emission from a broad distribution of velocities is a unique signature 
of C-type shocks.

The magnetic field strength, parameterised by $\beta$, is  crucial for determining the shock structure and the distribution
of molecules behind the shock front.
Observations indicate $\beta\sim 1$ in interstellar clouds; however,  there is
significant uncertainty and true spread in $\beta$, because
only the radial component is measured, and the conversion to the full magnetic field strength is challenging \citep{crutcher2012}. 
For $\beta=1$, shocks faster than 21 \kms\ are J-type.
Using the ram pressure of a supernova blast wave at $10^4$ yr after explosion, this corresponds to 
shocks into gas with density $n_0<10^4$ cm$^{-3}$. 
Zeeman observations of OH masers from supernova shocks indicate very high field strengths $\sim 1$ mG \citep{brogan2000} in the compressed, post-shock gas.
A shock into gas with pre-shock density $10^4$ cm$^{-3}$ and compression factor 10 
(which matches the density required in the masing region), the inferred $\beta\sim 3$.
We ran elevated magnetic field models with a modest increase to $\beta=5$.
The supernova blast wave ram pressure is expected to drive 80 \kms\ shocks into $n_0=10^3$ cm$^{-3}$ gas, which would be C-type if $\beta >4$,
so in order to model C-type shocks at lower densities, we modeled shocks into $n_0\leq 10^4$ cm$^{-3}$ with both $\beta=1$ and $\beta=5$.
Another consideration is that 
only the component of the magnetic field vector parallel
to the shock affects the 1-D models, 
so the effective $\beta$ is scaled by the cosine of the angle between the magnetic field 
and shock velocity vectors. 
The shocks into high-density ($n> 10^4$) gas with $\beta=1$ are C-type, so to model the case with
a decreased or misaligned magnetic field,
we generated models for $\beta=0.1$ and $\beta=1$.

\begin{figure}
    \centering
    \includegraphics[width=\linewidth]{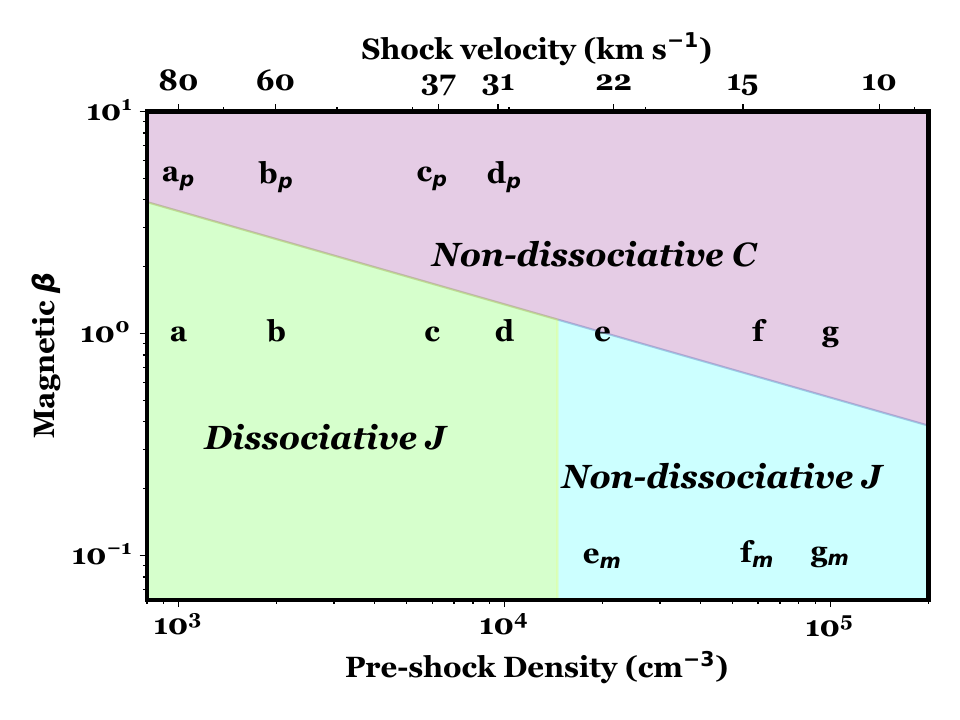}
    \caption{Phase space of shock parameters. The three critical regimes (non-dissociative C-type, dissociative J-type, and Non-dissociative J-type) are labeled (and shaded purple, green, and cyan, respectively). The input parameters for the shocks in Table~\ref{emisstab} are indicated by their corresponding letter designations.
    Model (e) falls near the C- to J-type boundary and was found to be C-type. This phase space is in accord with similar ones from \citep{godard2019,kristensen2023} diagrams, 
    }
    \label{cjdiag}
\end{figure}

The phase space of critical model parameters---pre-shock density ($n_0$), shock velocity ($V_s$), and magnetic field ($\beta$),
and the pre-shock density, shock velocity, and magnetic $\beta$ for the set of fiducial models is listed in Table~\ref{emisstab}.
This phase space is for a fixed ram pressure equal to that of  $10^4$-yr
old supernova remnants appropriate fro IC~443 and 3C~391, so the combination of $n_0$ and $V_s$ is tied together
(see Paper 1).
The boundary between C-type and J-type shocks (discussed in the previous paragraph) is indicated
in Figure~\ref{cjdiag}.
Table~\ref{emisstab} lists
the suite of models in order of increasing $n_0$ from (a) through (g).
Models with a nominal magnetic field ($\beta=1$) have no subscript, and the models with elevated and decreased 
$\beta$  have subscripts $_p$ or $_m$, respectively.
The locations of all these models are shown in Figure~\ref{cjdiag}.
An important distinction among the J-type shocks is whether H$_2$ molecules are destroyed.
H$_2$ molecules can survive shocks slower than 25 \kms\ \citep{tielens2005}. 
For our model phase space, the high density ($n_0>10^4$ cm$^{-3}$), low-$\beta$ (oblique or reduced magnetic field) are  non-dissociative, J-type shocks. 

\def\extraextinction{
The external radiation field and extinction between the shock front and the radiation field are important 
factors in determining the post-shock evolution. We set the incident radiation to the pre-shock gas 
equal to that of the 
solar neighborhood interstellar radiation field. 
We then modeled the pre-shock conditions as a PDR with 1.6 magnitudes of extinction, to establish the pre-shock 
gas conditions. These parameters approximate what may apply at the outside of a molecular clump in IC 443 and 3C~391, where
the atomic and low-density molecular gas are shielding the pre-shock gas by at least 1.6 magnitudes as 
evidenced by the extensive molecular gas seen in CO(1-0) imaging \citep{lee2012,wilner1998}.
}

\begin{figure*}
\includegraphics[width=3in]{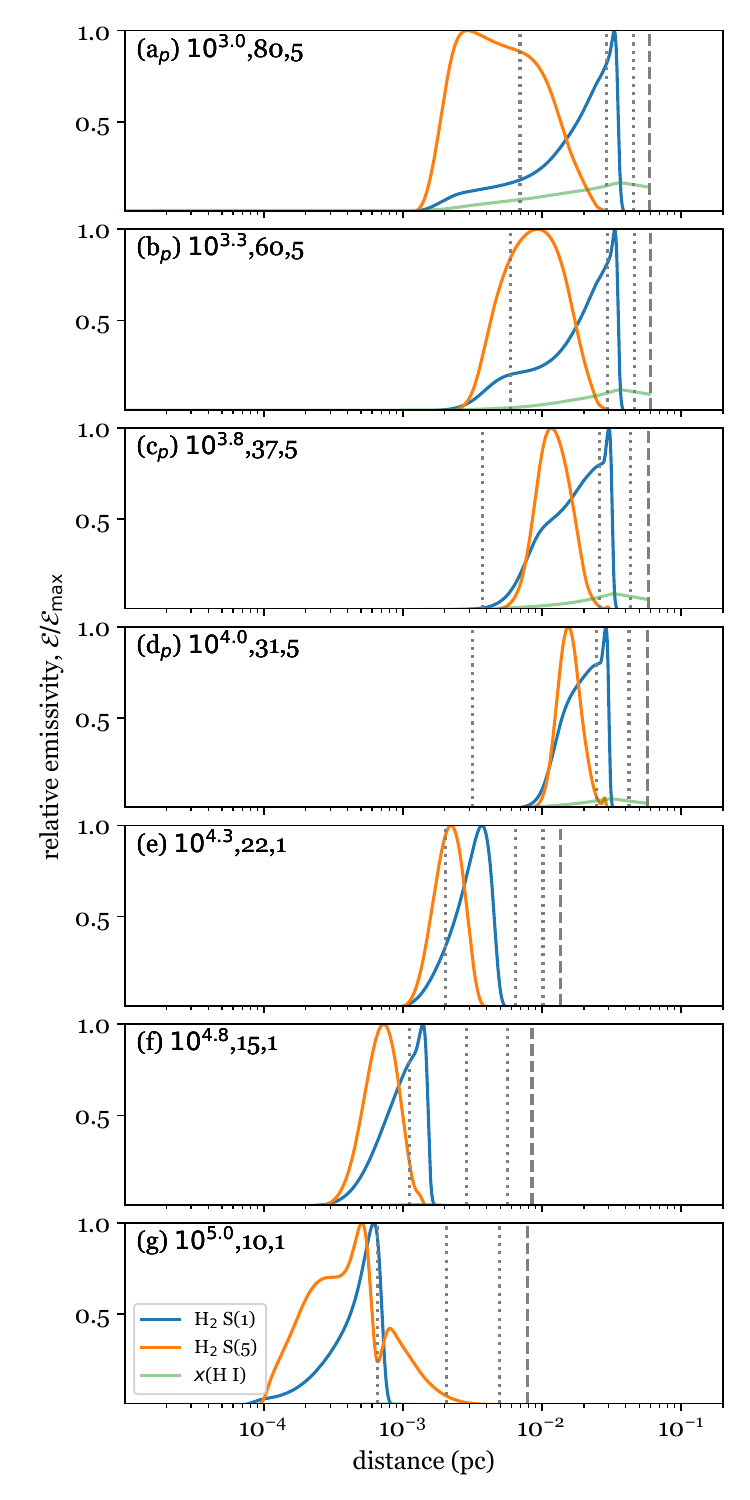}
\includegraphics[width=3in]{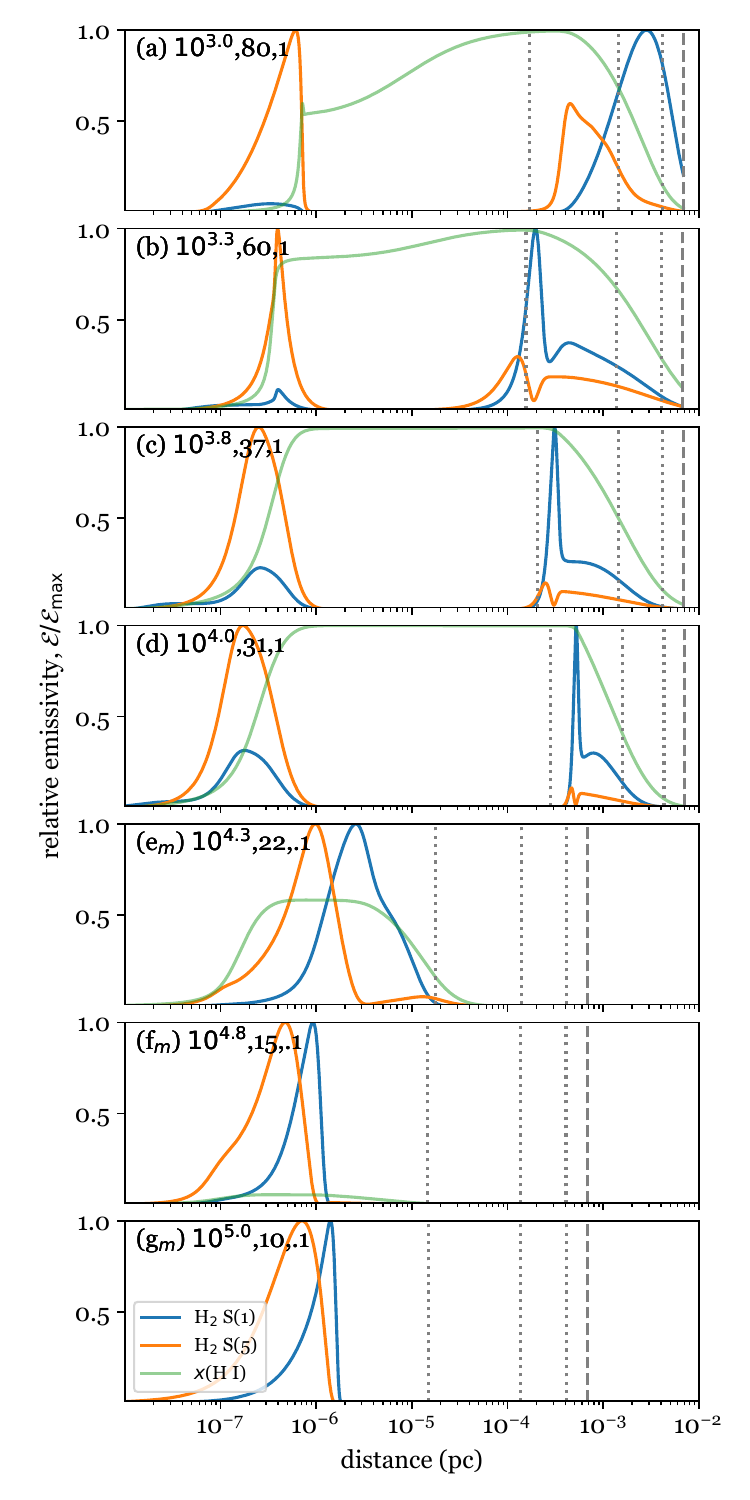}
\caption{Modeled emissivity of the H$_2$ S(1) and S(5) lines
as a function of distance behind the model shocks listed in Table~\ref{emisstab}.
The emissivities of the H$_2$ lines are normalized to the peak value for each modelled transition.
To show where molecules are dissociated, the fractional abundance of atomic H is shown in each panel, without normalization (so a value of 1 means complete
dissociation of H$_2$).
Each panel is labeled with its shock designation from
Table~\ref{emisstab} and  abbreviated 
pre-shock density (cm$^{-3}$), shock velocity (km~s$^{-1}$), and
magnetic parameter $\beta$.
The \textbf{left} column shows the C-type shock models, which become progressively more compact as the pre-shock density increases.
The \textbf{right} column shows the J-type shock models.
The dissociative J-type models (a, b, c, d) have a thin region just behind the shock front $2\times 10^{-7}$ pc), followed by a region  where molecules are destroyed, then 
followed by a region of reformed molecules (around $10^{-3}$ pc). 
Note the two columns have different horizontal scales.
The vertical dashed line shows the shock duration of 5000 yr, and
vertical dotted lines show how far the shock developed in (from right to left)
3000, 1000, and 100 yr.
}\label{fig:emissivity}
\end{figure*}

\begin{deluxetable*}{ccccccccccccccc}
\tablecolumns{13}
\tablecaption{Distance and temperature of emitting region \tnm{a}\label{emisstab}} 
\tablehead{
\multicolumn{4}{c}{Model} && \multicolumn{2}{c}{H$_2$ S(1) } && \multicolumn{2}{c}{H$_2$ S(5) } && \multicolumn{2}{c}{H$_2$O 1$_{10}$-$1_{01}$ }\\
\cline{1-4}
\cline{6-7}
\cline{9-10}
\cline{12-13}
\colhead{} & \colhead{density} & \colhead{shock speed} & \colhead{$\beta$}   &&
\colhead{$\langle T\rangle$} & \colhead{$\ell_{\rm emiss}$} &&
\colhead{$\langle T\rangle$} & \colhead{$\ell_{\rm emiss}$} &&
\colhead{$\langle T\rangle$} & \colhead{$\ell_{\rm emiss}$}  \\
&\colhead{(cm$^{-3}$)} & \colhead{(\kms)} &  &&
\colhead{(K)} & \colhead{(mpc)} &&
\colhead{(K)} & \colhead{(mpc)} &&
\colhead{(K)} & \colhead{(mpc)} 
}
\startdata
\cutinhead{C-type}
(a$_p$) & $10^3$     & 80  & 5 && 1400 & 24   && 4000 & 9      && 1100 & 26\\
(b$_p$) & $10^{3.3}$ & 60  & 5 && 1300 & 23   && 2600 & 12     && 1100 & 26 \\
(c$_p$) & $10^{3.8}$ & 37  & 5 &&  920 & 19   && 1300 & 13     &&  820 & 31\\
(d$_p$)     & $10^4$ & 30  & 5 &&  770 & 19   &&  940 & 17     &&  700 & 29\\
(e) & $10^{4.3}$     & 22  & 1 &&  870 & 3.4  && 1490 & 2.3    &&  410 & 5.5\\
(f) & $10^{4.8}$     & 15  & 1 &&  700 & 1.1  && 1020 & 0.8    &&  500 & 1.4\\
(g) & $10^5$         & 10  & 1 &&  520 & 0.5  &&  350 & 0.9    &&  330 & 0.7\\
\cutinhead{J-type}
(a)     & $10^3$     & 80  & 1  && 420  & 3.4   &&  860 & 1.5   &&  410 &  11\\
(b)     & $10^{3.3}$ & 60  & 1  && 150  & 2.0   &&  210 & 2.2   &&   94 &  11\\
(c) & $10^{3.8}$     & 37  & 1  && 190  & 1.5   &&  290 & 1.7   &&   97 &  11\\
(d) & $10^4$         & 30  & 1  && 150  & 1.3   &&  200 & 1.7   &&  120 &   7\\
(e$_m$) & $10^{4.3}$ & 22  & 0.1&& 480  & 0.006 && 1940 & 0.008 &&  220 &  0.05\\
(f$_m$) & $10^{4.8}$ & 15  & 0.1&& 1100 & 0.001 && 1040 & 0.002 &&  290 &  0.03\\
(g$_m$) & $10^5$     & 10  & 0.1&& 1720 & 0.001 && 1710 & 0.002 &&  490 &  0.04\\
\enddata
\tablenotetext{a}{For each spectral line $\ell_{\rm emiss}$ and $T_{\rm emiss}$ are the 
emissivity-weighted distance and temperature of the emitting region behind the shock front. 
}
\end{deluxetable*}

\begin{figure}
\includegraphics[width=.45\textwidth]{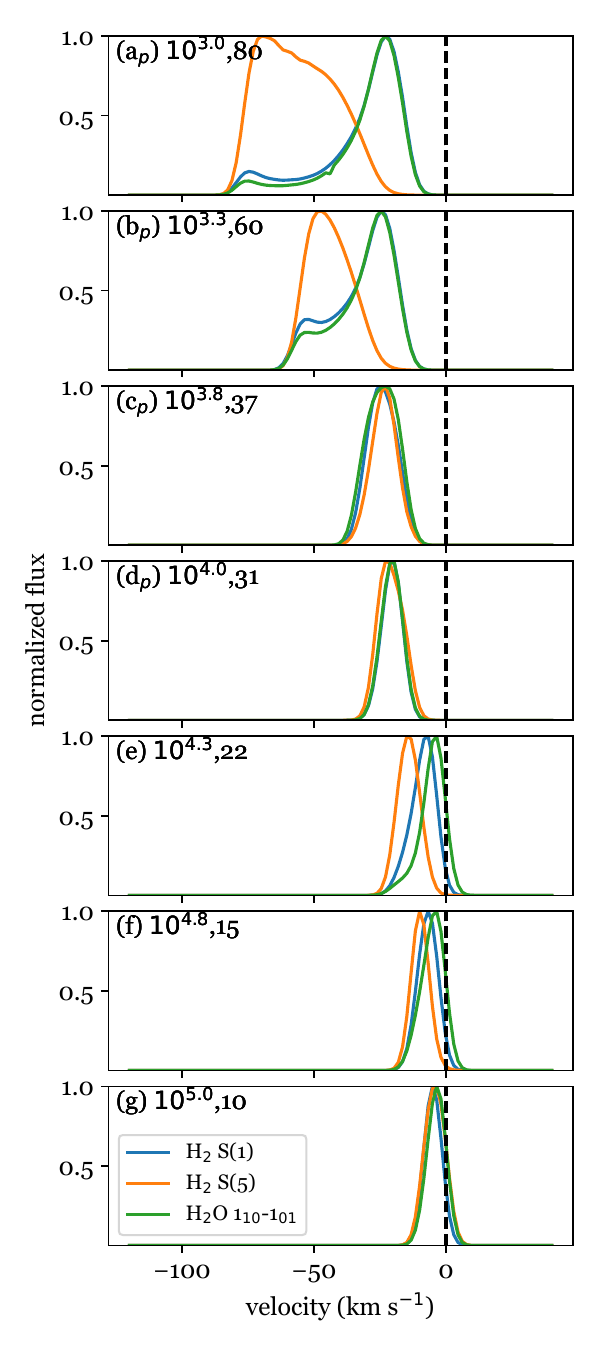}
\caption{Modeled spectra of  H$_2$ emission in the S(1) and S(5) pure rotational lines,
and the H$_2$O $1_{10}$-$1_{01}$ line, for 
C-type MHD shocks.
Each panel is labeled with its shock designation from
Table~\ref{emisstab} and  abbreviated 
pre-shock density (cm$^{-3}$), shock velocity (km~s$^{-1}$), and
magnetic parameter $\beta$.
All spectra are normalized by their peak values, which are reported in Table~\ref{modtab}. 
Velocities are in the  rest frame of the observer.
The dashed line at 0 \kms\ indicates the pre-shocked gas
velocity, and the modeled shock is propagating toward the observer
(along the line of sight). 
Unlike Figure~\ref{fig:emissivity}, the J-type models are not shown, because they are all
simple Gaussians centered near zero.
}\label{fig:modspectra}
\end{figure}

\begin{deluxetable*}{ccccccccccccccc}
\tablecolumns{15}
\tablecaption{Velocity distributions from shock models\tnm{a}\label{modtab}} 
\tablehead{
\colhead{} & \multicolumn{4}{c}{H$_2$ S(1) } && \multicolumn{4}{c}{H$_2$ S(5) } && 
\multicolumn{4}{c}{H$_2$O 1$_{10}$-$1_{01}$ }\\
\cline{2-5}
\cline{7-10}
\cline{12-15}
\colhead{model} &     
\colhead{$\langle v \rangle^\tnm{b}$} & \colhead{FWHM$^\tnm{c}$} & \colhead{$I_{\rm peak}^\tnm{d}$} & \colhead{$I_{\rm tot}^\tnm{e}$}&&
\colhead{$\langle v \rangle^\tnm{b}$} & \colhead{FWHM$^\tnm{c}$} & \colhead{$I_{\rm peak}^\tnm{d}$} & \colhead{$I_{\rm tot}^\tnm{e}$}&&
\colhead{$\langle v \rangle^\tnm{b}$} & \colhead{FWHM$^\tnm{c}$} & \colhead{$I_{\rm peak}^\tnm{d}$} & \colhead{$I_{\rm tot}^\tnm{e}$}
}
\startdata
\sidehead{C-type}
(a$_p$)    &  -21.5 &     34 &   0.43 & 0.23&&  -49.5 &     40 &   1.24 & 3.9&&  -17.6 &     26 &   2.34 & 0.034\\
(b$_p$)    &  -22.9 &     33 &   0.64 & 0.34&&  -41.6 &     27 &   2.63 & 5&&  -18.3 &     26 &   5.05 & 0.075\\
(c$_p$)    &  -21.2 &     21 &   1.20 & 0.65&&  -27.3 &     13 &   2.18 & 1.9&&  -17.7 &     19 &  15.58 & 0.23\\
(d$_p$)    &  -19.8 &     17 &   1.52 & 0.78&&  -22.8 &     11 &   1.01 & 0.73&&  -16.6 &     15 &  26.39 & 0.38\\
(e)        &  -10.2 &     14 &   0.71 & 0.26&&  -15.4 &     11 &   5.50 & 4&&   -6.3 &     13 &  71.48 & 0.68\\
(f)        &   -8.2 &     11 &   1.00 & 0.31&&  -10.7 &      8 &   3.42 & 1.9&&   -5.0 &     10 & 215.62 & 1.8\\
(g)        &   -5.9 &      8 &   0.84 & 0.19&&   -7.1 &      8 &   0.20 & 0.11&&   -3.9 &      9 & 144.20 & 1.1\\
\sidehead{J-type} 
(a)        &   -1.3 &      8 &   0.01 & 0.0012&&   -1.3 &      8 &   0.06 & 0.033&&   -1.3 &      8 &   0.01 & 6.4e-05\\
(b)        &   -1.3 &      8 &   0.01 & 0.0015&&   -1.3 &      8 &   0.11 & 0.057&&   -1.3 &      8 &   0.00 & 1.5e-05\\
(c)        &   -1.3 &      7 &   0.01 & 0.0022&&   -1.4 &      8 &   0.24 & 0.12&&   -1.3 &      8 &   0.01 & 6.9e-05\\
(d)        &   -1.3 &      7 &   0.02 & 0.0033&&   -1.3 &      8 &   0.32 & 0.16&&   -1.3 &      8 &   0.02 & 0.00015\\
(e$_m$)    &   -0.2 &      7 &   0.04 & 0.0081&&   -0.5 &      7 &   0.37 & 0.17&&   -0.2 &      7 & 359.92 & 2.2\\
(f$_m$)    &   -0.3 &      7 &   0.06 & 0.011&&   -0.3 &      7 &   0.66 & 0.3&&   -0.2 &      7 & 256.97 & 1.5\\
(g$_m$)    &   -0.7 &      7 &   0.02 & 0.0042&&   -0.7 &      7 &   0.76 & 0.35&&   -0.4 &      7 &  95.10 & 0.57
\enddata
\begin{raggedright}
\tablenotetext{a}{For shocks directed along the line of sight. Models are labeled as in Table~\ref{emisstab} and 
Figs~\ref{cjdiag}, \ref{fig:emissivity}, and~\ref{fig:modspectra}.}
\tablenotetext{b}{Mean velocity of the emission line profile, in \kms.}
\tablenotetext{c}{Full width at half-maximum intensity of the emission profile, in \kms.}
\tablenotetext{d}{Peak of the spectral intensity in observational units of Jy~arcsec$^{-2}$}
\tablenotetext{e}{Integrated line surface brightness, in units of $10^{-3}$ erg~s$^{-1}$~cm$^{-2}$~sr$^{-1}$}
\end{raggedright}
\end{deluxetable*}

\def\olddata{
(a$_p$) & -21.0 & 40 & 7.3  & 4.5   && -48.7 & 33 & 21  & 48   && -17.6 & 36 & 1.7 & 0.03\\
(b$_p$) & -21.6 & 28 & 13   & 7.9   && -26.9 & 19 & 20  & 27   && -18.2 & 27 & 4.0 & 0.07\\
(c$_p$) & -24.2 & 13 & 14   & 5.1   && -23.5 & 12 & 0.2 & 0.2  && -17.7 & 14 & 4.6 & 0.06 \\
(d$_p$) & -20.6 &  9 &  3   & 0.7   && -20.9 & 11 & 0.1 & 0.1  && -20.6 & 10 & 6.4 & 0.05\\
(e)     & -8.8  & 11 & 10   & 2.9   && -14.1 & 10 & 43  & 2.9  && -5.5  & 12 & 130 & 1.3\\
(f)     & -7.0 &   9 & 10   & 2.6   && -9.7  &  8 & 23  & 1.2  && -5.2  & 10 & 202 & 1.9\\
(g)     & -5.0 &   8 &  1   & 0.2   && -4.2  &  9 & 0.2 & 0.1  && -3.7  &  8 & 168 & 1.2\\
\sidehead{J-type} 
(a)     & -1.3 &  8 &   9   & 2.7   && -1.4 &  7 & 2.5  & 1.3  && -1.3  &  8 & 1.0  & 0.01\\
(b)     & -1.3 &  8 & 0.1   & 0.02  && -1.3 &  8 & 1.2  & 0.7  && -1.3  &  8 & 0    & 0\\
(c)     & -1.3 &  7 & 0.1   & 0.03  && -1.4 &  7 & 2.7  & 1.4  && -1.3  &  8 & 0.01 & 0\\
(d)     & -1.4 &  7 & 0.2   & 0.04  && -1.4 &  7 & 4.2  & 2.0  && -1.3  &  8 & 0.02 & 0\\
(e$_m$) & -0.2 &  7 & 0.7   & 0.01  && -0.5 &  7 & 3.0  & 2,2  && -0.2  &  7 & 364 & 2.2 \\
(f$_m$) & -0.6 &  7 & 0.2   & 0.01  && -0.6 &  7 & 9.6  & 3.6  && -0.2  &  7 & 173 & 1.6\\
(g$_m$) & -1.0 &  7 & 0.07  & 0.01  && -1.0 &  7 & 4.6  & 4.4  && -0.3  &  7 & 67 & 0.6

}


The distinctions between the different shock types are evident in the H$_2$ distribution and emissivity.
Figure~\ref{fig:emissivity} shows the emissivity of the H$_2$ S(1) and S(5) lines for the models listed
in  Table~\ref{emisstab}.
The atomic H density, which is 
effectively emissivity of the 21-cm atomic H hyperfine line,
is included in Figure~\ref{fig:emissivity} to show where H$_2$ molecules are dissociated.
The distinction between the J and C type shocks is  clear when comparing the left- and right-hand sides of the Figure.
The dissociative J-type shocks (models a,b,c, and d) 
have two regions of H$_2$ emission, with atomic H filling the gap in between.
A hot emitting region close ($3\times 10^{-7}$ pc)
behind the shock front is where the molecules are being destroyed; it is spatially so thin that it does not contributes more
than 1\% of the integrated emission for the models considered here. (But some species, like CH$+$, only
exist in this hot region because they are quickly chemically converted in the colder region where
molecules reform.)
The non-dissociative J-type shocks (models e$_m$, f$_m$, and g$_m$) look more like C-type shocks, because molecules 
survive. However, the lack of strong magnetic field means neutrals and ions stay together, which affects the shock structure.
Comparing to the stronger magnetic field C-type shocks into the same densities (models e vs. e$_m$, f vs. f$_m$, and g vs. g$_m$), the J-type shocks are much 
more compact. The H$_2$ emitting regions of C-type shocks are of order $10^{-3}$ pc wide, while the non-dissociative J-type equivalents are only of order $10^{-6}$ pc wide.

Table~\ref{emisstab} summarizes the distance of the emitting regions behind the shock fronts, and their temperatures.
The average emitting distance was calculated as follows:
\begin{equation}
    \ell_{\rm emiss} \equiv \frac{\int \ell \mathcal{E}(\ell) {\rm d}\ell} {\int  \mathcal{E}(\ell) {\rm d}\ell},
\end{equation}
where $\mathcal{E}(\ell)$ is the emissivity (erg~cm$^{-3}$)
of a spectral line versus distance $\ell$ behind the shock.
The C-type shocks become progressively more compact for regions of higher density, and the H$_2$ emissivity for the
S(5) line is closer to the shock than that of the S(1) line. This is intuitively obvious because the H$_2$ molecules
survive the entire shock, emitting as they cool, and the S(5) line is from a higher energy level tracing
more highly-excited gas.
Note that if one were to use the ratio of observed brightness of two spectral lines to
determine the gas temperature, the result could be misleading because the emission in the two lines do not arise from
the same location. The H$_2$ S(1) line arises from gas cooler than where the S(5) line arises, and the
H$_2$O $1_{10}-1_{01}$ line arises from gas that is cooler still, in keeping with their respective upper level energies
above ground state.

\begin{figure*}
\includegraphics[width=.33\textwidth]{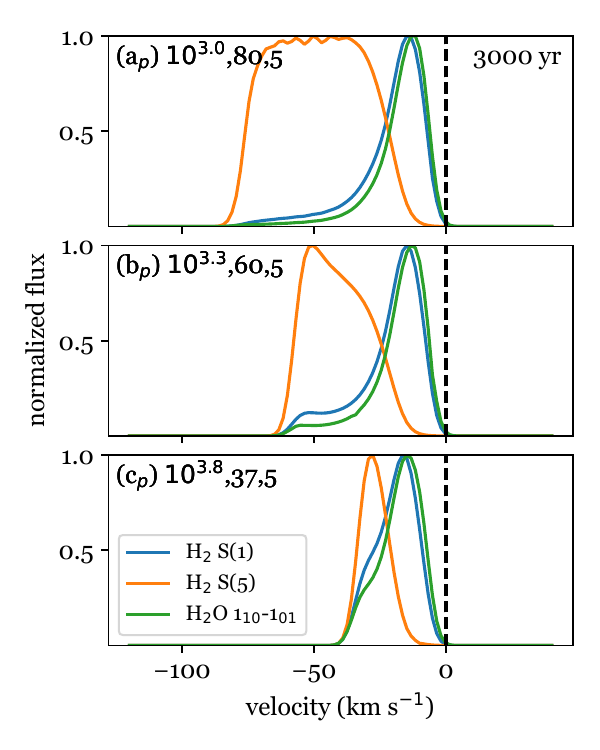}
\includegraphics[width=.33\textwidth]{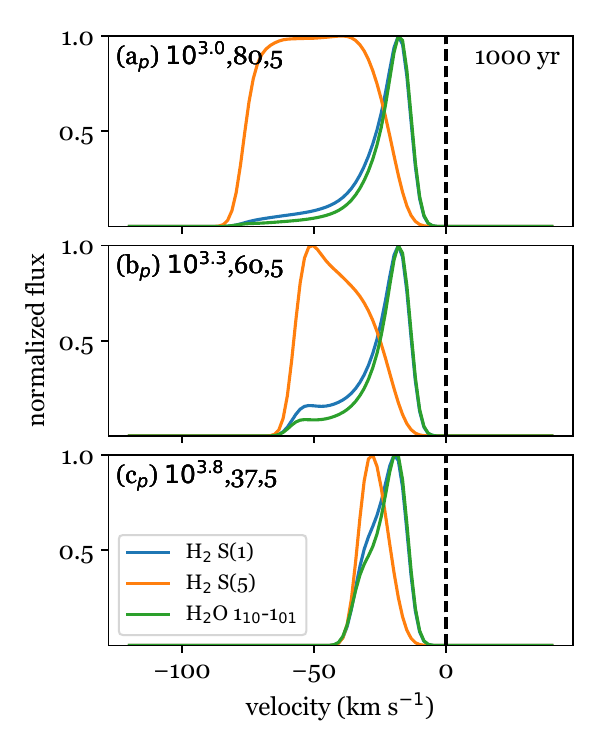}
\includegraphics[width=.33\textwidth]{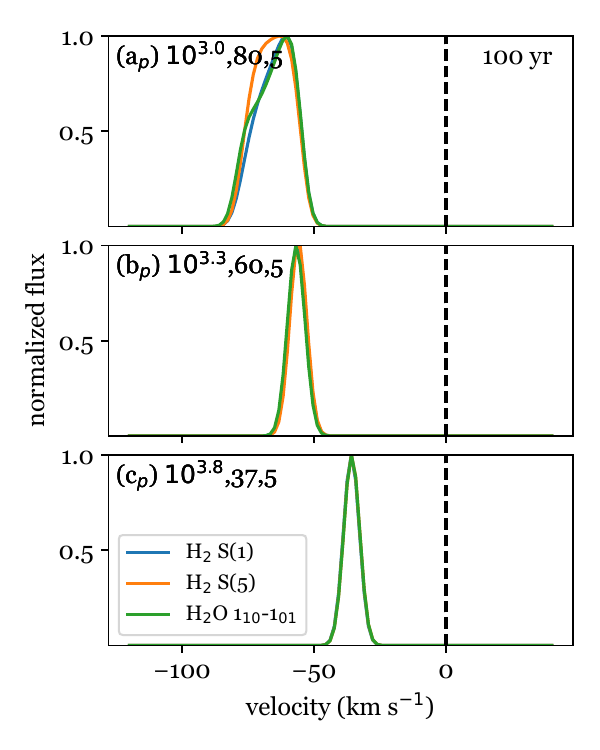}
\caption{Same as Fig.~\ref{fig:modspectra}, but for only the three fastest C-type shocks a$_p$, b$_p$, and c$_p$,
at three different ages of 3000 yr (left), 1000 yr (center), and 100 yr (right).
The velocity distribution skews away from zero if the shocks are younger, because 
the accelerated gas closer to the shock becomes a larger fraction of the emission.
}\label{fig:modage}
\end{figure*}

When predicting the emission from a short-lived phenomenon like a supernova remnant, the duration 
of the shock must be considered.
The supernova remnants are only $\sim 10^4$ yr old, which is not much more than the cooling time of the gas.
Using Figure 30.1 of \citep{draine2011}, the cooling time for gas with density $10^3$ cm$^{-3}$
is of order $10^3$ yr.
If the shock front took a fraction $f$ of the total time $10^4$ yr since the explosion to reach the shocked clump, then
the shock duration is $(1-f)\times 10^4$ yr. 
In both IC~443 and 3C~391, the shocked molecular clumps are close to the edges of their SNR, and they span a 
range of projected distances any explosion site, suggesting $1>f\simgt .5$, which means shock duration from 0 to 5,000 yr.
Inspecting the temperature profiles of the shocked gas, the gas cools within 3000 yr for all models.
For the shocks into gas with $n_0> 10^4$ cm$^{-3}$,  the shocks are essentially stationary 
because cooling times are short. 
Vertical lines in Figure~\ref{fig:emissivity}  illustrate how far the shock front 
develops in a clump  if it arrived  100, 1000, 3000, and 5000 yr ago.
Figure~\ref{fig:modage} shows the emitted spectra for different ages.
The gas in very young shocks has not decelerated, so that the velocity distributions skew further from zero.
For the youngest shock (100 yr), the emission arises from just behind the shock front,
at velocity $V_s$. None of the observations discussed in this paper are likely to be such young shocks, because they
all have emission spanning $V_0$ to $V_S$.
But the accelerated clump IC 443:B2 \citep{reach2019} may be such a case, where the blast wave has only
recently reached the clump, perhaps because it is farther from the explosion site.
In the remainder of this paper, we use the models for a shock duration of 3000 yr, which is plausible in terms of 
the supernova remnant age and widespread locations of the shocked clumps.

Synthetic emission spectra were calculated for each model using the emissivity versus distance from Figure~\ref{fig:emissivity}
and a velocity distribution at each distance that is centered on the neutral gas velocity and has a local width that is
a combination of sound speed at that distance and a fixed turbulent dispersion
(3 km~s$^{-1}$).
Figure~\ref{fig:modspectra} shows the predicted spectra from the three different shock models in each of
the H$_2$ S(1) and S(5) lines, for the case where the shock propagates along the line of sight and towards the observer. The rest velocity of the  pre-shock gas is at 0 \kms. 
If the shock is propagating perpendicular to the line of sight, the emission
lines  all center on 0 \kms, and there is no further diagnostic information from the spectral shape.
For intermediate inclination angles, the centroid velocity of the shocked emission shifts approximately as the cosine of
the inclination angle (if inclination 0 is defined as propagating parallel to the line of sight).

Synthetic H$_2$O $1_{10}-1_{01}$ spectra were calculated using the 
velocity, density, temperature, and H$_2$O abundance at each distance
behind the shocks. The excitation of the upper level was calculated with the collision rates from \citep{green1993} and
radiative rates from \citet{pickett2003}. Collisional rates were multiplied by 1.35 to include
collisions with He.
At each distance behind the shock, the collisional rates were evaluated by interpolating between
tabulated temperatures for the downward rates. The upward transition rates, $\gamma_{lu}$, were precisely calculated from the downward rates, $\gamma_{lu}$, based
on detailed collisional balance.
\def\extra{
\begin{equation}
\gamma_{lu} = \gamma_{ul} \frac{g_u}{g_l} e^{-E_{ul}/kT}
\end{equation}
where $u$ and $l$ denote the upper and lower levels,  $E_ul$ is the energy difference between the levels, $k$ is the Boltzmann constant,  $T$ is the neutral gas temperature. }
This step prevents numerical error that arise
from using tabulated upward and downward rates. 
An important caveat to the predicted H$_2$O model line profiles is that we assumed the emission is optically thin. A treatment of optically thick emission is beyond our present scope, though we note it has been 
attempted for CO \citep{fuda2023}
The resulting synthetic H$_2$O $1_{10}-1_{01}$ spectra are very similar to those of the H$_2$ S(1) line, so
we did not include them in the figures. The H$_2$O emitting region thickness, line flux and peak intensity for each model is included 
in Tables~\ref{emisstab} and \ref{modtab}.

\section{Comparing Observations and Models}

\subsection{Pre-shock conditions}

First we must establish the rest velocity of the pre-shock gas and the direction of the shock along the line of sight.
The ambient, pre-shock molecular gas into which the IC~443 blast wave is propagating has been mapped in low-J CO emission,
and it arises mostly at radial velocities
between -6 to -1 \kms\ with respect to the 
galactic neighborhood local standard of rest (LSR)
\citep{lee2012}. 
For 3C~391, the parent molecular cloud is at 90--110 \kms from the CO (1--0) mapping
\citep{wilner1998}, and the ambient gas absorbs the shocked emission in the CO(2-1) spectra and
produces narrow CS(2-1) emission at 105 \kms \citep{reach1999}, which we adopt as the rest velocity.
The velocity range of the shocked emission relative to the
ambient gas, and the relative velocity spreads of the S(1) and
S(5) lines, allow us to determine the orientation of the shocks,
which is required before comparing to the models. 
If the shock were propagating perpendicular to the line of sight, the velocity gradient behind the shock would be masked and
all lines would be centered on the pre-shock velocity.
This appears to be approximately the case for IC~443~G and 3C~391, which agrees with the presence of significant self-absorption
in the molecular lines from the lowest energy levels or ground state.

If the shock is propagating along the line of sight, then the shocked gas will be displaced in velocity
as predicted by the synthetic spectra in Figure~\ref{fig:modspectra} with centroids in Table~\ref{modtab}.
The H$_2$ emission from IC~443~B and IC~443~C is significantly offset from the pre-shock velocity, indicating those
shocks are along the line of sight.
The shocked gas (Fig.~\ref{fig:spectra}) is at more negative velocities than the ambient gas for both supernova remnants, meaning the shocks are propagating toward us along the line of sight.

\subsection{Velocity range}

For IC~443~B, the H$_2$ and H$_2$O emission spans the entire range from 0 to -80 \kms. 
As explained above, we can infer from this that $V_0\sim 0$ \kms, $V_s\sim 80$ \kms, and
the shock is moving toward the observer.
Using only this shock speed, and the IC~443 blast wave pressure,
the pre-shock density is of order $10^3$ cm$^{-3}$. 
Therefore, the initial guess is that models (a) and (a$_p$) may apply to IC~443~B. 
In Paper 1, we fit models to the S(5) line profiles and found the
best-fitting pre-shock density $2\times 10^3$ cm$^{-3}$ and
shock speed 60 \kms, which is similar.
The C-type shock with elevated magnetic field, model (a$_p$) provides a reasonably good description of the observed range of velocities. The corresponding J-type model, (a),
is centered near the velocity of the pre-shock gas and has a width of only 8 \kms, which
is not at all what was observed. No combination of fully dissociating J-type models can
explain the observed velocity distribution.
Model (a$_p$) also correctly predicts that the H$_2$ S(1) line will be concentrated at velocities lower than the S(5) line. 
\def\extra{
To make this comparison quantitative, consider the velocities at half-maximum brightness.
For the S(1) line, the observed full width at half-maximum (FWHM) is 21 \kms, while the modeled FWHM is 17 \kms.
For the S(5) line, the observed full width at half-maximum (FWHM) is 63 \kms, while the modeled FWHM is 41 \kms\.
The modeled S(5) line is somewhat narrower, primarily because it 
lacks some of the observed emission on the lower-velocity end.
This deficit could be made up by including some shocks into higher-density gas such as seen for the other clumps, which we turn to next. As mentioned above in describing the observed
2-D spectra (Fig~\ref{fig:s5image}), the spatial profiles
also support the idea that some of the lower-velocity S(5) emission is spatially distinct from the wider wing.
}

For IC~443~C, the observed S(1) and S(5) lines are relatively similar
in shape, with a two-component structure.
There is a bright component at -8 \kms\ with FWHM 15 \kms, and a 
wider component centered at -26 \kms\ with FWHM 28 \kms.
The wide velocity spread indicates the shocks are propagating
at least partially along the line of sight (again, toward the
observer). 
The observed FWHM is smaller than for IC~443~B, so a lower-speed shock into higher-density gas applies to IC~443~C.
In Paper 1, the S(5) line profile was well-fitted
by a combination of a 60 \kms\ shock into density $2\times 10^3$ cm$^{-3}$
and a 31 \kms\ shock into a higher-density of $10^4$ cm$^{-3}$,
similar to models (b$_p$) and (c$_p$). 
The S(1) line profile shape is more similar to the the higher-density shock model.
Referring to Figure~\ref{fig:modspectra}(b), the models predict
S(1) and S(5) lines with similar widths and centers, as
observed.

For IC~443~G, the observed S(1) and S(5) lines are again relatively similar and have a two-component structure, but the narrow component is much brighter than the wider component.
The narrow components of the S(1) and S(5) lines have FWHM of 10 and 14 \kms, which are  similar to the narrow
components of IC~443~C.
The deep H$_2$O absorption  provides a clue that the pre-shock velocity is -4 \kms, 
and the observed emission centroids
of the S(1) and S(5) lines are -6 and -8 \kms, respectively. 
The small shift between the shocked emission and the pre-shocked
gas indicates that the shock is slightly inclined from the line of sight.
In Paper 1, a viewing angle of $65^\circ$ was inferred.
The small shift in velocity between the S(1) and S(5) lines is
in the sense predicted by the shock models for this viewing angle.

\subsection{Line brightness and S(1)/S(5) ratio}

The absolute brightness of the observed shocks can be a diagnostic of the shock, as the H$_2$ lines are a primary coolant \citep{flower2015}.
\def\extra{
Observed peak line brightnesses of the S(5) lines are in the range 0.4--2.0 Jy~arcsec$^{-2}$.
The predicted S(5) peak brightness, $I_{peak}$,
from Table~\ref{modtab}, is large enough to explain the observed
emission for all of the J-type models and all but the highest-density (model g) C-type models.
The observed S(1) peak brightnesses are in the range 0.4--1.8 Jy~arcsec$^{-2}$,
which can be explained only by the C-type models.
The line ratio S(1)/S(5) is in the range 0.5--1.5, which is predicted by only by C-type models a$_p$, b$_p$
and possibly f, but by none of the J-type models.
}
The brightness scales approximately as the the secant of the inclination with respect to the line of sight,
limited by the size of the shocked clump relative to the emitting region thickness. 
The increased brightness for shocks perpendicular to the line of sight is partially
canceled out by the finite angular resolution of the observations.
From Table~\ref{emisstab}, the emitting region behind a shock into $10^3$ cm$^{-3}$ gas,
at the distance of IC~443, would subtend $2.1''$ in the S(1) line and $3.8''$ in the S(5) line,
which is comparable to the SOFIA angular resolution of the observations \citep[$3.7''$ at the wavelengths of both the S(1) and S(5) lines:][]{temi2018}; thus the beam filling factor should be of order unity for
shocks into $10^3$ cm$^{-3}$ gas, whatever the inclination relative to the line of sight.
Shocks perpendicular to the line of sight are also limited by the pre-shock clump size and by the filling
factor of the shock in the beam. 
The size of dense cores and filaments in molecular clouds is  of order $\ell_{\rm clump}=0.1$ pc
\citep{andre2014}.
The angular resolution of the SOFIA observations is $3.7''$ at the wavelengths of both the S(1) and S(5) lines \citep{temi2018}, which is $\ell_{\rm beam}=0.036$ pc at the distance of IC~443.
The line brightness can be increased for shocks inclined with respect to the line of sight
by as much as $\ell_{\rm clump}/\ell_{\rm beam}=2.8$.
The observed brightnesses are all lower than the predicted ones, by factors of $\sim 10$, taking
all these factors into account. That the brightnesses are this close to the predictions is a testament to
their validity, as the modeled physics includes MHD, thermodynamics, and chemistry of a complicated
and likely non-homogeneous medium.

The ratio of brightness of the S(1) to S(5) lines is a diagnostic of the shocks independent of 
both the beam filling factor and the inclination angle. 
For IC~443~B, the ratio of S(1)/S(5) is 1.5 at the peak of S(1), but it decreases to $<0.2$ at 
velocities $<-50$ \kms, where S(1) is not detected.
For IC~443~C, the ratio S(1)/S(5)  is 0.3 in the narrow  component (and 0.5 tentative wide component).
For IC~443~G, the ratio S(1)/S(5) is 0.9.
These are in the observational units of Jy~arcsec$^{-2}$ for peaks of the line profiles.
For the models in Table~\ref{modtab}, the S(1)/S(5) ratio in the same units is 0.3 for model a$_p$ and 0.7 for
model b$_p$, and those models are the best single fits to the line ratios for the broad components of the
observed spectral lines. The narrow components have significantly higher S(1), and they arise from
cooler gas such as from shocks into higher-density gas or gas that has a long time to cool.

\begin{figure}
\includegraphics[width=.45\textwidth]{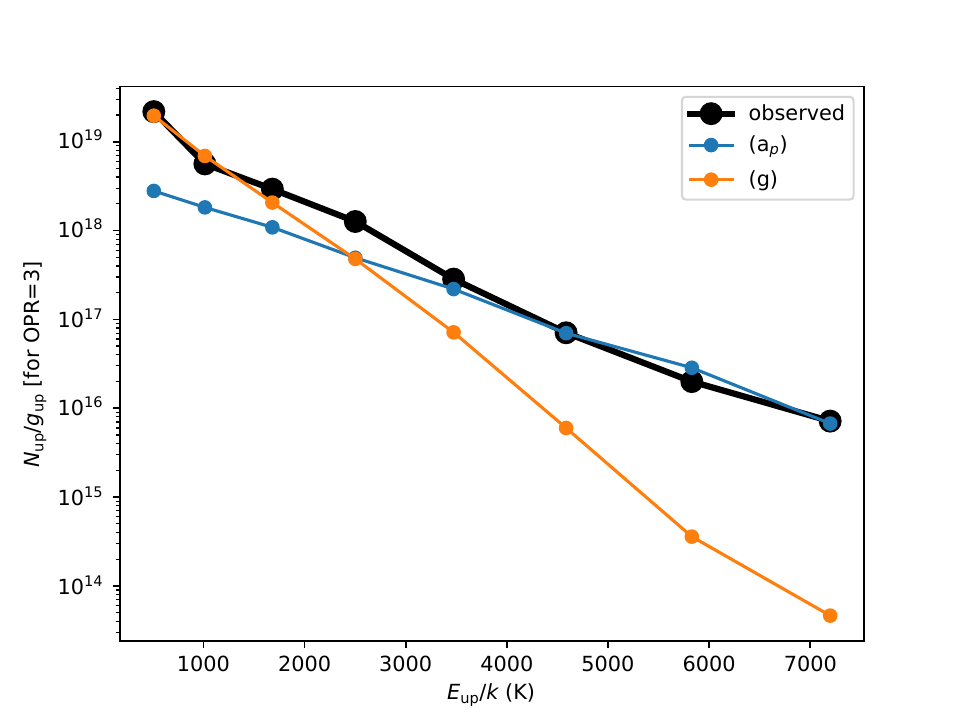}
\includegraphics[width=.45\textwidth]{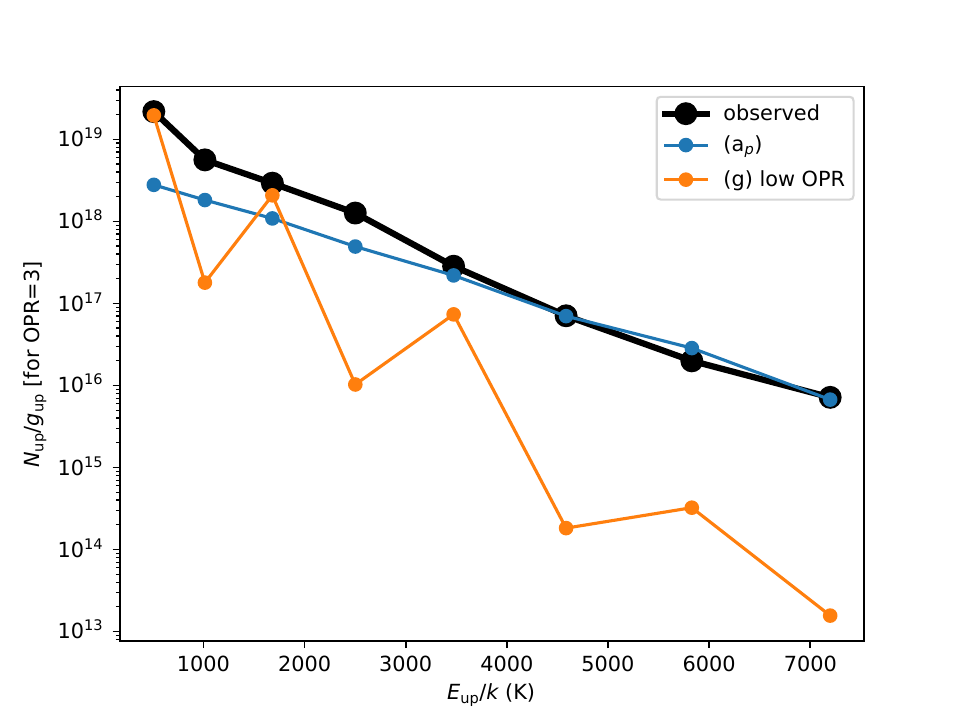}
\caption{Column density divided by statistical weight for the upper levels
of the pure rotational S(0) through S(6) lines toward IC~443~C. 
The horizontal axis is the energy (expressed as a temperature) of the upper level.
The observed values are in black.
The predictions from 
shocks into moderate-density gas (combination of models (b$_p$) and (c$_p$) from Table~\ref{emisstab}), normalized to match the observed brightness of the S(5) line, are in blue.
The predictions from a shock into higher-density gas (model (g) from Table~\ref{modtab}), normalized to match the observed brightness of the S(0) line, is in orange.
The upper panel is for a shock into gas with statistical equilibrium populations of H$_2$ (OPR=3).
The lower panel is for a shock into gas that has been cold and undisturbed (OPR=0.006). 
}\label{fig:excitation}
\end{figure}

A broader view of the H$_2$ emission is needed, so we include
{\it Spitzer} \citep{werner2004,houck2004} observations of H$_2$ lines S(0) through S(7) from prior work on
these targets \citep{neufeld2007,yuan2011,dellova2024}. 
The observed flux is directly proportional to the column density in the upper level.
Figure~\ref{fig:excitation} shows the column densities versus energy of the upper level,
where column densities were divided by their statistical weight assuming an
ortho-to-para H$_2$ abundance ratio of 3.

Model (a$_p$), which is required to explain the width of the S(5) line, can  predict
the column densities of  the upper levels of the S(4) through S(7) lines.
But that model, when normalized to match the
observed brightness of the S(5) line, under-predicts the observed flux of the
S(0) line by a factor of 8.
Shocks into denser gas are much more prolific in the lower-J lines, so
we added a second model (g): a 10 \kms\ shock into gas with
density $10^5$ cm$^{-3}$.
Model (g) was normalized to match 90\% of the observed flux
in the S(0) line. The combination of these two shock models is able to match the H$_2$ excitation diagram over
the full set of S(0) through S(7) lines.
The relative contributions model (a$_p$):model (g) for the SOFIA-observed S(1) lines
is predicted to be 80\%:20\% for the S(1) line and 8\%:92\% for the S(5) line. 
This explains why there is a broad component, fainter in S(1) than in S(5), and a narrow component, 
relatively brighter in S(1) in the line profiles.

The H$_2$ ortho-to-para ratio (OPR) for the dense, pre-shock gas 
can influence the OPR in the post-shock gas, potentially affecting our
conclusion that a shock into dense gas is required.
For shocks faster than about 20 \kms, the OPR is completely reset to 3 
in the shock front, so the initial OPR only slightly changes the shock structure.
This is because the peak temperature behind a faster shock 
is high enough, for a long enough time,
that the ortho-to-para ratio of the shocked gas is reset to the
high-temperature equilibrium value, and the gas retains that value
for longer than the $10^4$ yr age of the supernova remnant.
\citet{neufeld2024} showed that the ortho-to-para ratio 
for the S(4) through S(7) lines is near the high-temperature limit, 
in agreement with this prediction of the models regardless of their initial OPR.
But for a slow, 10 \kms\ shock, the OPR can remain far
from equilibrium, if the pre-shock gas was out of equilibrium. 
Therefore, we compared model (g) (which had OPR=3) with alternatives:
a very-low initial OPR=0.006 
(which arises from $10^4$ cm$^{-3}$ gas cooling for 60 Myr) and a moderate OPR=1.
For the very-low initial OPR model, the shock increases the OPR to only 0.15, and
the emitted spectrum from the shocked gas has the distinctive, zig-zag signature of  
odd versus even levels shown in the lower panel of Figure~\ref{fig:excitation}.
For the moderate initial OPR=1 model, the shock increases the OPR slightly, to 1.2, 
and the gas remains near OPR=1 throughout the age of the supernova remnant.
The observations show only a slight zig-zag excitation diagram from S(0) to S(3),
indicating the pre-shock gas had $1<$OPR$<3$.
The observed brightness
of the ortho-H$_2$ S(1) and S(3) lines, compared to the para-H$_2$ S(0) and S(2) lines,
constrains the OPR of the shocked gas to be in the range $2\pm 0.5$. 
This result is similar to what was found from fitting
 other supernova remnant H$_2$ excitation
diagrams \citep{hewitt2009}.
In conclusion, the  amount of dense, shocked gas can be confidently inferred from
the model fits, even using the ortho-H$_2$ S(1) line, because the ortho-to-para ratio
is of order unity.

\subsection{Model fit to velocity distribution}

\begin{figure}
\includegraphics[width=.5\textwidth]{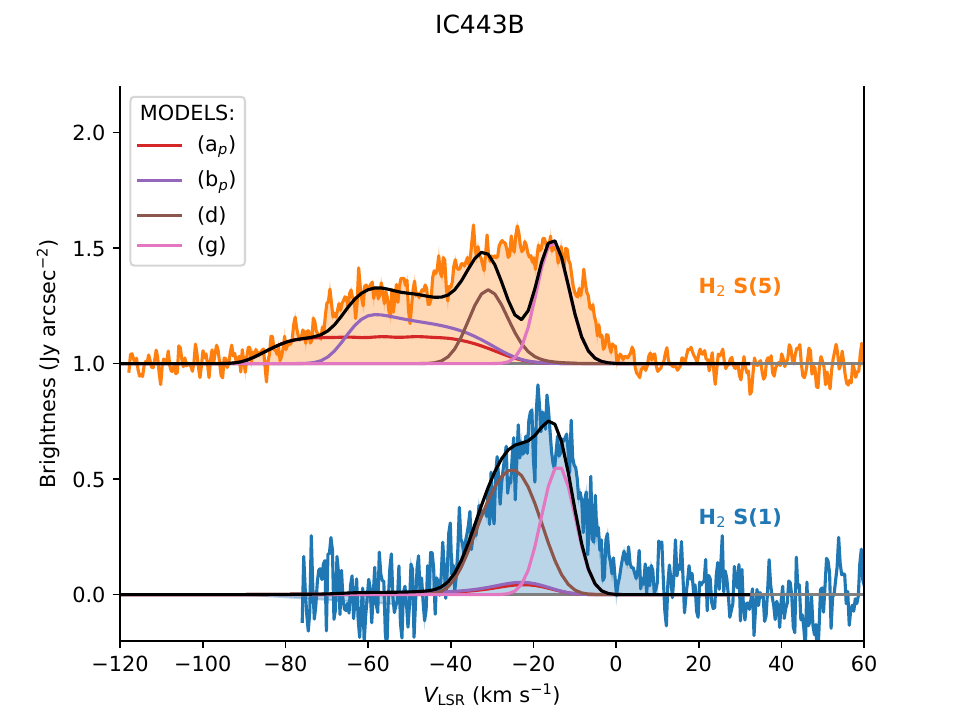}
\includegraphics[width=.5\textwidth]{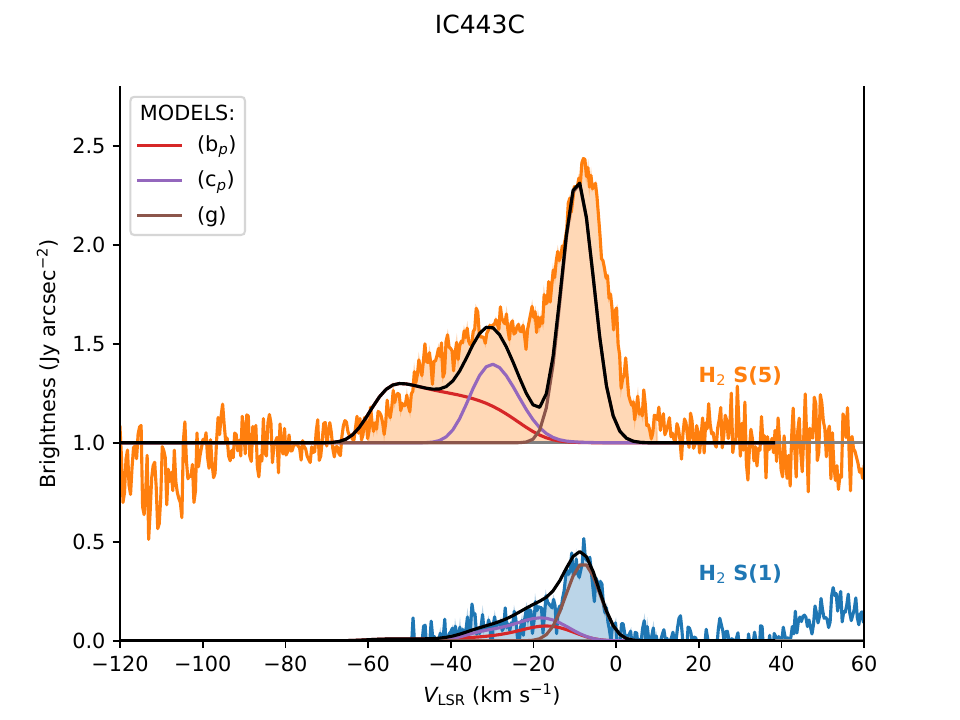}
\caption{Observed H$_2$ S(5) and S(1) spectra of IC~443~B (top) and IC~443~C (bottom) with scaled model spectra.
The models are labeled with their name from Table~\ref{emisstab}.
The sum of  models (black curve) matches the overall velocity distribution. 
The bulk of the modeled S(5) flux is from shocks into gas 
with moderate densities, while the  S(1) line is mostly from
shocks into high-density gas, although all models contribute to both lines.
}\label{fig:ic443b}
\end{figure}

The velocity distribution is the best indicator of the shock speed and post-shock gas kinematics.
To minimize projection effects, and to maximize radial velocity spread,
we consider in some detail the shocked clump with the broadest and most asymmetric velocity distribution
(which is most likely to be a shock along the line of sight): IC~443~B.
Figure~\ref{fig:ic443b} shows how a linear combination of  models from Table~\ref{emisstab} compares to the observed spectra.
Because we only used a sparse model grid, the combined models have `bumps' corresponding to the peak velocities of
each model used in the input. Overall, the models can explain the velocity distribution well in
both the S(1) and S(5) lines. The model combination was built as follows.
First, the lowest density (but enhanced magnetic) model, (a$_p$), was required in order to explain the 
highest-velocity observed gas. 
Then, a high density model, (g), was required to explain the brightness of the S(1) line.
Two intermediate models were then added to span the broad, continuous velocity distribution.
The linear combination that matches the observed spectra had filling factors 
for models (a$_p$), (b$_p$), (d), and (g) of .1, .1, .3, and .7, respectively. Thus the dense clump was a 
significant presence in the beam, and the lower-density gas 
fills a small fraction of the beam.

\def\extra{
We approximated a continuous distribution of shock models by interpolating between the models at $10^3$,
$10^{3.3}$, and $10^{3.8}$ cm$^{-3}$ with  100 logarithmic spaced densities, and weighted the models by
a power law proportional to the inverse squared logarithm of the density.
The sum of these models fits the high-velocity part of the S(5) profile very well, and it fits most
of the S(1) emission, but it
misses gas at velocities around -20 \kms\ that is present both in the H$_2$ S(1) and S(5) lines. 
The unexplained emission could arise from relatively low-velocity shocks into high-density gas, 
but the centroid velocity is too far shifted from the ambient gas for this to work.
Instead, we added an obliquely viewed shock model, specifically model (d) from Table~\ref{modtab} 
viewed at inclination $70^\circ$.
This combination of properties is difficult to explain with oblique shocks, but it
can be partially explained by shocks into high-density gas. The model with density $10^4$ cm$^{-3}$ provides
no significant help in fitting the line profile (because it was already near the range of the continuous distribution).
Instead, shocks into $10^5$ cm$^{-3}$ gas can explain part of the emission. In Figure~\ref{fig:ic443b}, the
high-density shock, added to the lower-density ones, improves the fit to the data, although not in
great detail. We infer that some higher-density gas is present in the cores, which is significant
in relating them to star-forming cores. The amount of shocked emission arising from the dense gas is relatively small.
}

\subsection{Dissociation of Molecular Hydrogen behind the shock}

The models that match the H$_2$ line profiles and excitation diagrams are C-type shocks,
where the H$_2$ survives the initial heating and emits as it cools. But such shocks 
only occur at high velocities when the magnetic field is both strong and oriented perpendicular to the shock velocity. For the same physical conditions, but a 
magnetic field oriented more parallel to the shock velocity, dissociative J-type shocks
are expected in the low-density gas, as delineated in Figure~\ref{cjdiag}.
We now test whether there is significant dissociation of H$_2$ by measuring
the amount and velocity distribution of atomic H behind the shock front.
Accelerated atomic gas has been noted before toward IC~443
\citep{braun1986,lee2008}.
For regions like the northeastern shell, this shocked atomic gas is
expected and associated with shocks into lower-density ($n<10$ cm$^{-3}$) gas.
The atomic gas in those regions 
arises from the recombined and cooling gas behind the very fast ($V_s>100$ \kms)
J-type shocks,
which are prominent
at optical wavelengths in H recombination and heavier-element fine-structure lines.
On the other hand, toward the southern molecular ridge that includes the shocked clumps studied in the present project, the pre-shock gas  has much higher density resulting in slower shocks.

To  measure the amount of atomic hydrogen that may be associated
with shocks into the molecular gas, we utilized the data from the  
Galactic Arecibo L-Band Feed Array (GALFA) \ion{H}{1} 21-cm 
line survey \citep{peek2011}.
The beam size of the former, great telescope at Arecibo was
$3'$, which samples the scales of the shocked clumps but 
does not resolve  dense cores.
IC~443 was  included in the Canadian Galactic Plane Survey, with slightly higher
angular resolution ($1'\times 3'$) showing similar results \citep{bakis2024}, 
and at even higher resolution ($40"$) with the Very Large Array \citep{lee2008}.
We use the single-dish spectra here for their high sensitivity.

\begin{figure}
\includegraphics[width=.5\textwidth]{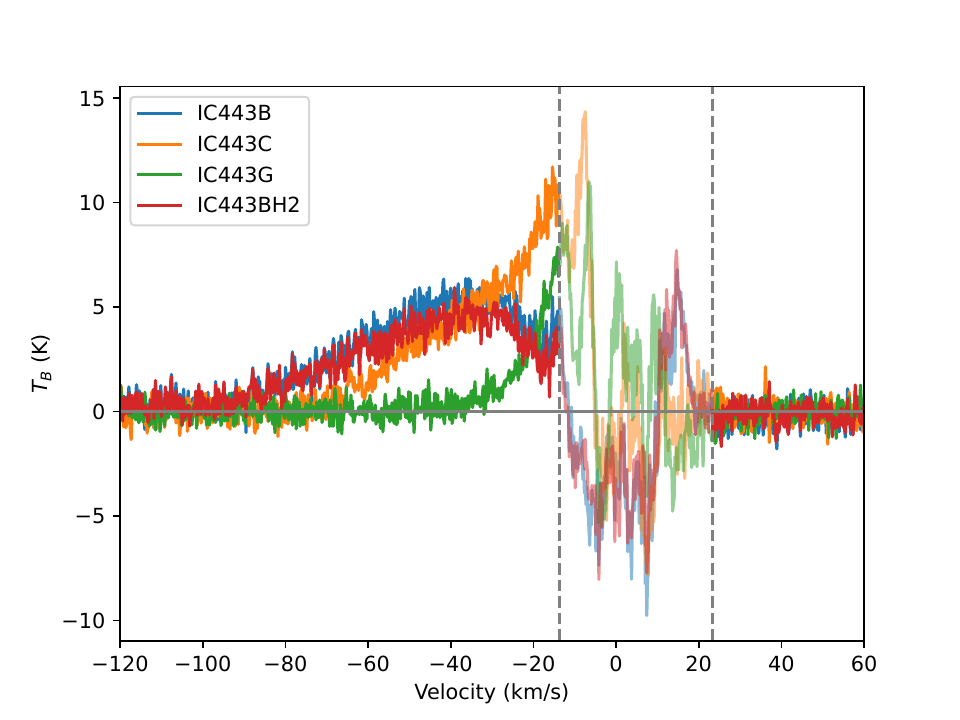}
\caption{21-cm line spectra toward the IC~443 shocked clumps B, C, and G.
Each spectrum is from a position centered on the clump, with nearby background subtracted.
At low velocities, indicated by the vertical grey dashed lines, the spectrum is strongly contaminated by absorption of
the supernova remnant synchrotron emission by cold foreground gas and should
be ignored.
}\label{fig:arecibo}
\end{figure}

Figure~\ref{fig:arecibo} shows the 21-cm line spectra
for the three clumps in IC~443.
The 21-cm line from each shocked clump was evaluated by locating the `on' spectrum
closest to the clump center, then subtracting the average of two `off' spectra
(for each clump) located 1.5 beamwidths away from the `on'.
For IC~443~B and C, 21-cm emission is present over a wide range of velocities. For IC~443~G, there is no such high-velocity atomic gas. 
The 21-cm line profiles (both width and centroid) are very similar to the H$_2$ and H$_2$O spectra, which strongly
suggests they arise from the same shocks.
The atomic H velocity distributions are most similar to model (a) from Table~\ref{emisstab}.
Thus the atomic gas appears to be associated with the shock into the moderate-density gas.
This makes physical sense, because a shock into the high-density gas with a speed of 10 \kms\ and the gas is never heated above about 5,000 K.
For a 60 \kms\ shock, on the other hand, the peak temperature is above 15,000 K and destruction 
of H$_2$ molecules is possible.

The observed \ion{H}{1} column densities are well constrained, and they are only at the velocity
of the shocked gas. Interpolating over the region where the 21-cm line
could not be accurately measured (between the dashed lines in Figure~\ref{fig:arecibo}), 
the  column density of atomic H in the shocked gas is $4\times 10^{20}$ cm$^{-2}$ for IC~443~B and C, and $<1\times 10^{20}$ cm$^{-2}$ for IC~443~G. 
In comparison,  models (a$_p$), (b$_p$), and (c$_p$) predict an \ion{H}{1} column density of $2\times 10^{20}$,
due to partial dissociation of H$_2$ in the shocked gas
as shown by the green curves in Figure~\ref{fig:emissivity}.
The amount of \ion{H}{1} is smaller by a factor of at least 10 for the other C-type shocks.
The J-type shocks, on the other hand, are completely dissociated for models (a) through (d), and
they generate \ion{H}{1} column densities at least 10 times higher than observed. 
The low-magnetic-field, high-density, J-type shocks (e$_m$), (f$_m$), and (g$_m$) are
non-dissociative but have much lower \ion{H}{1} column density than observed.

The \ion{H}{1} column densities are thus associated with the same shocks 
into moderate-density gas,
both in terms of the atomic column density and the velocity distribution,
as the ones that explain the H$_2$ and H$_2$O emission. 
For the range of velocities corresponding to molecules partially destroyed in model ($a_p$), the
spatial distribution of the 21-cm emission is shown in Figure~7 of \citet{lee2008} in blue and for models ($b_p$ and $c_p$) in green. The shape of the molecular ridge is recognizable in that 21-cm image, with blue peaks 
at IC~443~B and IC~443~C where the shocks are largely along the line of sight. (IC~443~G is not
noticeable due to the shock being perpendicular to the line of sight.) The 21-cm emission
shows a similar velocity and spatial distribution to the H$_2$. 

To assess the relative amount of atomic and molecular gas directly from the observations of the shocked clumps,
we estimate
the fraction of the hydrogen nuclei 
that are in atomic gas:
\begin{equation}
    f_{\rm atomic} = \frac{N({\rm H I})}{2 N({\rm H}_2)+N({\rm H I})}.\label{eq:fdiss}
\end{equation}
The atomic H column density is directly traced by the observed line profile, because the gas is in 
the optically thin limit where the integrated brightness temperature is proportional to column density. 
The H$_2$ column density was calculated from the observed brightnesses of the S(0) through S(7) lines, adding in
the column density in the ground state calculated from the excitation temperature of the lowest to levels each of ortho- and para-H$_2$. 
The column densities in levels $J=2$ through $J=9$ are directly observed, and
the excitation diagram is extrapolated to measure the ground ortho- ($J=1$) and para- ($J=0$) abundances.
For the shocks into dense gas, most H$_2$ molecules are in the $J=2$ and $J=3$ levels;
only 28\% of the H$_2$ is in the lower $J=0$ and $J=1$ levels. 
For the shocks into moderate-density gas, most H$_2$ molecules are in the $J=3$ and $J=4$ levels; only 8\% are in the $J=0$ and $J=1$ levels.
The dissociation fractions estimated this way are about 30\% if we consider all of the H$_2$ associated with the same shock as the \ion{H}{1}. If we consider only the H$_2$ from shocks into moderate-density gas, the dissociation fraction is about 60\%.
The shocks into dense gas could have as little as zero dissociation of H$_2$.
These observational results are in agreement with the shock models used above, which 
predict the 21-cm line to arise from partially-dissociating C-type shocks into moderate-density gas.

\def\extra{
Assuming the gas was entirely molecular before passage of the shock, then Equation~\ref{eq:fdiss} is 
related to the fraction of molecules dissociated by the shock and not yet reformed.
For a long-lived shock, the column density of cold, reformed molecules grows approximately linearly with time.
Therefore, it is possible to use Equation~\ref{eq:fdiss} to estimate the age of the shock. 
We find that the atomic fractions are very high, which requires the shocks to be young.
For example, for a 37 \kms\ shock into $n_0=6\times 10^3$ cm$^{-3}$, the \ion{H}{1} column
density is $3\times 10^{20}$ cm$^{-2}$, and the H$_2$ column density is 
$3\times 10^{20} t_{\rm shock,3}$ cm$^{-2}$ where $t_{\rm shock,3}$ is the shock duration in thousands of years.
Using Equation~\ref{eq:fdiss}, we can solve for a shock age of 1,000 years.
The 21-cm line observations thus lend support to the idea that the IC~443 molecular shocks are young,
as independently inferred from the populations of the H$_2$ levels \citep[e.g.][]{cesarsky1999,reach2019}.
}

\subsection{Water abundance \label{sec:h2oabsorption}}

In the pre-shock gas, the H$_2$O abundance can be directly estimated from the {\it Herschel} observations.
For IC~443~G, the H$_2$O $1_{10}-1_{01}$ spectrum shows a deep, narrow dip at -4 \kms. 
Because this transition is from the ground-state of ortho-H$_2$O, the absorption is due to cold H$_2$O along the line of sight, 
presumably in the pre-shock core. 
The emission profile from the warm, shocked gas is so much broader than the absorption dip that
we can approximate the emission as an effective continuum, fitting a straight line from the blue to red side of the absorption dip.
In this way, the dip is measured to be approximately $93\pm 2$\% deep.
The optical depth required to cause the dip is $2.6\pm 0.3$, which corresponds to a column density of ortho-H$_2$O of $1\times 10^{14}$ cm$^{-2}$.
From a multilevel analysis of the H$_2$ lines, the
H$_2$ column density of IC~443~G is approximately $10^{22}$ cm$^{-2}$
\citep{dellova2024}.
Thus the water abundance in the pre-shock core 
(including para-H$_2$O) is of order $1\times 10^{-8}$. 
This abundance is in the range found for  translucent
molecular clouds \citep{vandishoeck2013}.
In our models, the pre-shock H$_2$O abundance was calculated using a steady-state chemical network, which
led to initial abundances somewhat lower than the observed value. 

The models that match the H$_2$ lines predict H$_2$O lines that much brighter than observed, even after scaling by 
the filling factors found from the linear combinations of models to the observed H$_2$ lines. The velocity distribution
of the H$_2$O is also very broad, being more similar to the H$_2$ S(5) line despite that line being from a much higher energy level than 
that of the H$_2$O line. The model H$_2$O emission arises from a warm layer where the H$_2$O abundance determined from the
chemical network is tremendously enhanced to $x({\rm H}_2{\rm O})=3\times 10^{-4}$. For comparison, in the pre-shock
gas the model abundance is $4\times 10^{-10}$, and in the cooled post-shock gas it is $1\times 10^{-9}$. 
A likely explanation is that the H$_2$O $1_{10}-1_{01}$ line is optically thick, so that the our calculation
(which assumed they were optically thin) over-predicted the emergent intensity.
\citet{fuda2023} demonstrated how to calculate the emission of an optically thick CO line from a shock, but that calculation has
not been attempted for the H$_2$O ground-state line. To get the models to match the observations, we would require
an escape probability lower than $10^{-2}$, which is plausible but has not been verified.
Because the water chemistry is complicated and relates the gas and solid phases of the ISM, we do not
attempt to interpret the results further, and instead only use the H$_2$O velocity profile as an aid to
interpreting the shock dynamics.

\section{Conclusions}

\def\frompierre{
  First, I believe this is the finest discussion of molecular lines I've ever read in the literature, combining line shapes in HI, H2 and H2O (and a bit in CO, too, to pin point the rest frame of the pre-shocked gas). I feel you could boast about it a bit more in the conclusion and emphasise that combining high spectral spectral resolution instruments (SOFIA, Arecibo, ALMA, NOEMA) for multiple molecular tracers was a key in this analysis. Maybe even point out that complementary data on the excitatoin diagram (by JWST, why not) is useful.
  Second, the analysis itself, combining HI and H2 to evidence H2 dissociation in shocks, to my knowledge that was never done before ? Is this the first evidence for dissociation shocks in space ? Can it be reiterated elsewhere ?
}

The new observations and modeling in this paper reveal 
the velocity distribution, molecular dissociation, pre-shock gas conditions, and shock age
of supernova blast waves in dense molecular clouds.
The properties of shocks were inferred from high-spectral-resolution
observations of the H$_2$ pure rotational lines.
Using H$_2$ lines, as opposed to those from molecules of trace elements, 
is crucial because H$_2$ comprises the bulk of the gas mass,  the
low-lying lines of H$_2$ are the primary coolant of 
the shocked gas, and the H$_2$ abundance does not depend upon the complicated chemistry of the shocked gas.
Using the pure rotational lines is important because the excitation
behind the dense shocks excites H$_2$ primarily into the lower few rotational levels of the molecule,
though the first two vibrational levels are excited for shocks into the lowest densities considered in our models.
The velocity distribution of the molecular gas was directly observed using the H$_2$ S(1)
and S(5) lines. This type of observation was not previously possible because
the H$_2$ lines are mostly at wavelengths where the Earth's atmosphere is opaque.
SOFIA flew to high enough altitude to get above 99.9\% of the telluric absorption to allow
the line profiles to be measured. Prior measurements of the
H$_2$ velocity distribution used the ro-vibrational 1-0 S(1) line
\citep{rosado2007}, which  traces a high energy level but has
the advantage of high-angular-resolution two-dimensional imaging, revealing 
substructures (both spatial and spectral) down to the angular scale of our new observations.
There is limited potential to follow-up  our H$_2$ observations, apart from challenging 
ground-based observations of the H$_2$ 0-0 S(2) line, which may be possible with
a mid-infrared echelle spectrograph like TEXES \citep{lacy2002}; those observations
would trace the shock into high-density gas and would constrain the ortho-to-para
conversion of that gas. We found an ortho-to-para ratio for the shock into dense gas 
less than the equilibrium value of 3, which implies incomplete conversion of the cold, pre-shock 
gas from its initial value (if the molecular cloud had indeed been cold and undisturbed prior to the supernova).
The shock into higher-density gas can be readily observed in the near-infrared with the 1-0 S(1) line;
our models showed that the shock traced by the near-infrared lines will have ortho-to-para ratio of 3 regardless of the pre-shock value.
\def\extra{
A special case of a forbidden line of a heavier element that 
is produced by a wide 
range of shock types, as explained by \citep{hollenbachMoleculeFormationInfrared1989},
is the the [\ion{O}{1}] 63.2 $\mu$m line. For 3C~391, a high-resolution spectrum
shows the line is 100 \kms wide, indicating it is produced primarily by 
shocks into moderate-density gas. On that same line of sight, far-infrared H$_2$O, CO, and OH lines
are produced by shocks into denser gas 
\citep{reach2000a}. The new observations in this paper consider the velocity-resolved profiles 
of the S(1) and S(5) transitions of the most abundant molecule, H$_2$, tracing shocks into
high- and moderate-density gas, respectively.
}

Theoretical predictions from the Paris-Durham shock code reproduced many features of the observations, 
including the broad distribution of velocities for IC~443~B and IC~443~C that require C-type (non-dissociative) shocks largely along
the line of sight. For IC~443~G and 3C~391, the shocks are largely perpendicular to the line of sight, making the
radial velocities overlap with that of the preshock gas.
Indeed, for those two shocked clumps, their broad emission lines have superposed deep, narrow absorption features,
in H$_2$O ($1_{10}-1_{01}$) for IC~443~G and HCO$^+$(1-0) for 3C~391,
due to cold, dense, unshocked gas from the part of the dense core between the shock front and the observer.

To explain the velocity distribution of the H$_2$ S(5) line, and the brightness 
of the S(4) through S(7) lines, preshock densities 
are in the range 1,000 to 6,000 cm$^{-3}$, into which the IC~443 blast wave drives shocks of speed 80 to 37 \kms, respectively. 
To explain the velocity distribution of the H$_2$ S(1) line, and the brightness of S(0) through S(3) lines, 
shocks into gas with density 10,000 to 100,000
cm$^{-3}$ gas are required.
Shocks into dense gas have been indirectly inferred before from mm-wave observations of the similar SNR W~44 \citep{seta2004,anderl2014}.
The need for a range of shock conditions arising from the same ram pressure is also seen
on much larger scales from a galaxy intruding into 
a group of galaxies \citep{appleton2006}, and is a natural
consequence of the interstellar medium having multi-scale structure (as opposed to clouds of constant density).

We used 21-cm line observations to measure the velocity distribution and abundance of
atomic H at the shock fronts, indicating the shocks dissociate less than half of the initial H$_2$.
This is one of the first times that partial dissociation of H$_2$ has been observed in the ISM \citep[cf. for similar SNR W~51C][]{koo1997}.
Complete dissociation (and partial ionization) occurs in shocks into lower-density ($< 10^2$ cm$^{-3}$) gas; however, the lack
of  optical emission lines indicates
that such shocks are not significant in the molecular-dominated part of supernova remnants like IC 443 or 3C 391 \citep{deng2023}. 
The Arecibo H~I observations  show that the atomic H has the same velocity distribution and peaks at the same
location as the shocked molecular gas. Future, high-angular-resolution observations of the 21-cm line with
observatories like the Square Kilometer Array \citep{braun2015} could study the dissociation behind 
shocks into molecular clouds in more detail.

The properties of the shocked clumps can be readily compared to those of star-forming dense cores.
The size scales of the shocked clumps B, C, and G in IC~443 are $\sim 0.4$ pc, with the mass of clump G 
being of order $100$ $M_\odot$ \citep{dickman1992,cosentino2022,dellova2024}.
The clumps are comparable to the widths of filaments in star-forming regions,
which have a typical diameter of 0.2 pc \citep{andre2014}.
The shocked clumps contain significant structure.
The spatial resolution of our observations is 0.03 pc, which is comparable to (but somewhat larger than)
the size of an individual protostar (Bonnor-Ebert condensation) \citep{myers2000}
with mass of order 1 $M_\odot$. 
These sizes are at the lower end of prestellar cores, for which lifetimes are estimated
to be 1 to 10 Myr \citep{jessop2000}. A likely scenario, then, is that the shocked
clumps were pre-existing dense cores of the parent molecular cloud of the progenitor star. 
The blast waves reached the dense cores about 8,000 years after the supernova explosion, and they
have been propagating into them for about 2,000 years.

The pre-existing dense cores were nearly massive enough to be self-gravitating,
if they had velocity dispersion less than 0.4 \kms (though observed line profiles
of the pre-shock molecular gas are typically larger than this). 
In the post-shock gas, where velocity gradients of several \kms\ are present, the
range of shock directions and amplification of turbulence are likely to keep them unbound or drive them from being gravitationally bound to being unbound.
The compression in the post-shock gas could be a trigger for star formation. For the
C-type shock models used in this paper (Table~\ref{emisstab}, top) that can explain the line profiles, the shock compression is predicted
to be a factor of 5 to 10. In contrast, for shocks into the same gas density but with
low magnetic field (J-type shocks; (Table~\ref{emisstab}, bottom), the compression is 
a factor of 30 to 170. The strong magnetic field, which is required to produce the 
observed, broad line profiles,  keeps the compression low and reduces the shock
front's potential to trigger star formation.
The shocks into IC~443 and 3C~391 molecular clumps are into dense cores that were
not imminently forming stars, and after the passage of the shock they may be even less likely to do so.

Future theoretical work on interaction of shock fronts with two- or three-dimensional structures with configurations
consistent with the best high-resolution observations would advance our understanding
of the relationship between supernova shocks and pre-existing molecular cores. 
Observational studies at higher angular resolution and including multiple energy levels, for example
in the infrared with JWST \citep{kristensen2023} or submillimeter with ALMA or NOEMA \citep{dellorusso2008},
will trace the shock fronts and provide a progressively more detailed view of how violent
events at the end of stellar evolution affect the potential of their parent molecular clouds
to form stars.

\begin{acknowledgments}
Based in part on observations made with the NASA/DLR Stratospheric Observatory for Infrared Astronomy (SOFIA). SOFIA was jointly operated by the Universities Space Research Association, Inc. (USRA), under NASA contract NNA17BF53C, and the Deutsches SOFIA Institut (DSI) under DLR contract 50 OK 0901 to the University of Stuttgart. 
Financial support for the observational and theoretical work in this article was provided by NASA through award \#07\_0007 issued by USRA to the Space Science Institute.
{\it Herschel} was an ESA space observatory with science instruments provided by European-led Principal Investigator consortia and with important participation from NASA.
\end{acknowledgments}

\facility{SOFIA,Herschel,Spitzer} 
\def\extra{
\software{
sofia\_redux \citep{clarkeReduxCommonInterface2015},
astropy \citep{astropycollaboration2022},
matplotlib \citep{hunterMatplotlib2DGraphics2007}
}
}
\def\apj{ApJ}
\bibliography{references}

\end{document}